\documentclass[american,aps,prl,twocolumn,longbibliography]{revtex4-2}
\usepackage[T1]{fontenc}
\usepackage[utf8]{inputenc}
\usepackage{amsmath}
\usepackage{amssymb}
\usepackage{graphicx}
\usepackage{wasysym}
\usepackage{babel}
\begin{document}
\title{Shape-space dynamics and geometric pattern formation in nonreciprocal
slender bodies}
\author{Balázs Németh}
\email{bn273@cam.ac.uk}

\author{Mohamed Warda}
\email{mrmaw2@cam.ac.uk}

\author{Ronojoy Adhikari}
\email{ra413@cam.ac.uk}

\affiliation{Department of Applied Mathematics and Theoretical Physics, Centre
for Mathematical Sciences, University of Cambridge, Wilberforce Road,
Cambridge CB3 0WA, United Kingdom}
\begin{abstract}
Nonreciprocal interactions in active solids violate action--reaction
symmetry and produce a net response to strain. Assuming invariance
under Euclidean symmetries, we derive a shape-space formulation for
the elastohydrodynamics of nonreciprocal slender bodies that separates
intrinsic deformation from rigid motion. The resulting nonlinear reaction--advection--diffusion
system represents a geometric flow whose activity-driven instabilities
generate steady, oscillatory, and chaotic patterns. These manifest
as rigid, swimming, and chaotic motion, linking nonreciprocal elastohydrodynamics
to geometric pattern formation and unifying recent observations in
slender active structures.
\end{abstract}
\maketitle
Slender elastic solids, spanning scales from eukaryotic flagella \citep{gray_propulsion_1955},
hair \citep{audoly_elasticity_2010}, and spaghetti \citep{audoly_fragmentation_2005}
to engineering structures \citep{baker_geometry_2025}, have long
served as paradigmatic systems in mechanics. Following the geometric
and nonlinear formulations of Euler, Kirchhoff, Timoshenko, and others
\citep{euler_methodus_1744,kirchhoff_uber_1859,timoshenko_lxvi_1921,ericksen_exact_1957,goriely_nonlinear_2000,antman_nonlinear_2004},
recent applications to plant growth \citep{moulton_multiscale_2020},
ciliary beating \citep{sartori_curvature_2016,chakrabarti_spontaneous_2019},
and locomotion \citep{rieser_geometric_2024,kaeser_individual_2025}
require accounting for two features: immersion in a viscous fluid,
where elastic and hydrodynamic forces compete in the overdamped elastohydrodynamic
regime \citep{lighthill_flagellar_1976,wiggins_flexive_1998,lauga_floppy_2007,butler_elastohydrodynamics_2026},
and sustained conversion of chemical free energy into work \citep{marchetti_hydrodynamics_2013},
generating internal forces and torques that do not derive from an
elastic potential.

Such systems may be viewed as active filaments \citep{chakrabarti_spontaneous_2019,makanga_instability_2026}
or slender active solids \citep{maitra_oriented_2019,brauns_active_2026},
with applications ranging from microscale transport\textbf{ }\citep{elgeti_physics_2015,parthasarathy_streaming-enhanced_2019}
to soft robotics \citep{ranzani_soft_2016,tekinalp_topology_2024}.
They have been realized experimentally using active colloids \citep{gopal_subramaniam_rigid_2024,chao_traveling_2025,wei_life-like_2026}
and robotic platforms \citep{brandenbourger_non-reciprocal_2019,chen_realization_2021,gelvan_hydrodynamic_2025,veenstra_adaptive_2025},
and modeled theoretically as discrete bead--spring chains \citep{jayaraman_autonomous_2012,winkler_physics_2020,prathyusha_emergent_2022}.
Within continuum theory, activity has primarily been introduced through
nonclassical constitutive laws or boundary conditions \citep{lough_self-buckling_2023}.
In odd elasticity \citep{scheibner_odd_2020,al-izzi_nonreciprocal_2026},
nonvariational stress--strain relations produce non-Hermitian phenomena
including unidirectional waves \citep{chen_realization_2021} and
self-organized swimming \citep{ishimoto_self-organized_2022}. Follower-force
models employ nonconservative boundary loads that generate buckling
and flapping instabilities \citep{de_canio_spontaneous_2017,ling_instability-driven_2018,clarke_bifurcations_2024,warda_elastohydrodynamic_2026}.
In flagellar motility, activity is represented through effective force
and torque densities acting on the elastic structure \citep{blum_biophysics_1979,mondal_internal_2020}.

Recent work has identified local sources of momentum \citep{poncet_when_2022}
and angular momentum \citep{nemeth_nonreciprocal_2026} in active
solids as generic consequences of nonreciprocal interactions, which
emerge upon coarse-graining microscopic degrees of freedom and violate
action--reaction symmetry. When the microscopic interactions are
invariant under Euclidean isometries, as in internally driven systems
or hydrodynamically interacting systems far from boundaries, these
sources depend only on strain \citep{nemeth_nonreciprocal_2026}.
Although the linear response of nonreciprocal solids is increasingly
understood \citep{maitra_oriented_2019,poncet_when_2022,nemeth_nonreciprocal_2026},
the role of geometric nonlinearities, which govern dynamics beyond
linear instability thresholds, remains unexplored.

Here we study the elastohydrodynamics of a geometrically nonlinear
slender solid driven by strain-dependent nonreciprocal force and torque
densities. Exploiting invariance under rigid motions, we formulate
the dynamics in shape space, decoupling intrinsic deformation from
physical-space configuration \citep{goldstein_nonlinear_1995}. The
resulting dynamics is a geometric flow described by a reaction--advection--diffusion
system for shape invariants \citep{cass_reaction-diffusion_2023}.
We show that this system undergoes pattern-forming instabilities and
bifurcations between fixed points, limit cycles, and chaos, which
reconstruct in physical space as rigid motion, swimming, and chaotic
dynamics. Nonreciprocal elastohydrodynamics thus becomes a problem
of geometric pattern formation in shape space, enabling direct application
of pattern-formation theory \citep{cross_pattern_1993,cross_pattern_2009}
and providing a unified framework for recent observations in slender
nonreciprocal active solids.

\textit{Elastohydrodynamics.} We consider a slender elastic solid
modeled as a thin uniform tube whose centerline is $\boldsymbol{r}\left(u,t\right)$
and to whose cross-section is attached an orthonormal moving frame
$\boldsymbol{e}_{1}\left(u,t\right),\boldsymbol{e}_{2}\left(u,t\right),\boldsymbol{e}_{3}\left(u,t\right)$,
where $u\in\left[0,L\right]$ is a Lagrangian parameter (not necessarily
arclength) and $t$ denotes time. Derivatives with respect to $u$
and $t$ are denoted by primes and dots, respectively. The kinematic
equations are 
\begin{equation}
\begin{aligned}\dot{\boldsymbol{r}} & =\boldsymbol{v}, & \dot{\boldsymbol{e}}_{i} & =\boldsymbol{\Omega}\times\boldsymbol{e}_{i},\\
\boldsymbol{r}' & =\boldsymbol{h}, & \boldsymbol{e}_{i}^{\prime} & =\boldsymbol{\Pi}\times\boldsymbol{e}_{i},
\end{aligned}
\label{eq:kinematics}
\end{equation}
where $\boldsymbol{v},\boldsymbol{\Omega}$ are the translational
and angular velocities and $\boldsymbol{h},\boldsymbol{\Pi}$ describe
deformation. The index $i=1,2,3$ and the vector $\boldsymbol{e}_{1}$
is normal to the cross-section. 

Euclidean isometries act as $\boldsymbol{r}\to\boldsymbol{R}\cdot\boldsymbol{r}+\boldsymbol{b}$,
$\boldsymbol{e}_{i}\to\boldsymbol{R}\cdot\boldsymbol{e}_{i}$ with
$\boldsymbol{R}\in\mathrm{SO}\left(3\right)$, $\boldsymbol{b}\in\mathbb{R}^{3}$.
Writing $\underline{A}=\left(A_{1},A_{2},A_{3}\right)$ for components
$A_{i}=\boldsymbol{e}_{i}\cdot\boldsymbol{A}$ in the moving frame,
the six-dimensional variables $E=\left(\underline{h},\underline{\Pi}\right)$
and $V=\left(\underline{v},\underline{\Omega}\right)$ are isometry-invariant
and describe the kinematics up to rigid motion. We refer to the set
of $E\in\mathbb{R}^{6}$ as the \textit{space of shapes}. The configuration
in physical space is reconstructed by integrating the kinematics from
$E$ and initial data.

In the elastohydrodynamic limit, the overdamped balance laws are 
\begin{equation}
\gamma^{T}\boldsymbol{v}=\boldsymbol{F}'+\boldsymbol{f},\quad\gamma^{R}\boldsymbol{\Omega}=\boldsymbol{M}'+\boldsymbol{h}\times\boldsymbol{F}+\boldsymbol{m},
\end{equation}
with internal stresses $\boldsymbol{F},\boldsymbol{M}$ and force
and torque densities $\boldsymbol{f},\boldsymbol{m}$. This assumes
that dissipative forces and torques are modeled by resistive force
theory, $\boldsymbol{f}^{D}=-\gamma^{T}\boldsymbol{v}$, $\boldsymbol{m}^{D}=-\gamma^{R}\boldsymbol{\Omega}$.
More accurate nonlocal kernels from slender-body theory \citep{garg_slender_2023}
should modify results quantitatively but not qualitatively. We assume
the tube is free to move so that $\boldsymbol{F}=\boldsymbol{M}=\boldsymbol{0}$
at its boundaries. Thus, overdamped motion is determined entirely
by instantaneous balance of momentum and angular momentum.

\textit{Shape-space dynamics.} Differentiating the kinematic equations
and equating mixed partial derivatives gives the compatibility conditions
\begin{equation}
\dot{\boldsymbol{h}}=\boldsymbol{v}^{\prime},\quad\dot{\boldsymbol{\Pi}}=\boldsymbol{\Omega}^{\prime}+\boldsymbol{\Pi}\times\boldsymbol{\Omega}.\label{eq:compatibility}
\end{equation}
Eliminating the velocities between the balance laws and the compatibility
conditions gives
\begin{align*}
\gamma^{T}\dot{\boldsymbol{h}} & =\boldsymbol{F}^{\prime\prime}+\boldsymbol{f}^{\prime},\\
\gamma^{R}\dot{\boldsymbol{\Pi}} & =\boldsymbol{M}^{\prime\prime}+\left(\boldsymbol{h}\times\boldsymbol{F}+\boldsymbol{m}\right)^{\prime}+\boldsymbol{\Pi}\times\left(\boldsymbol{M}'+\boldsymbol{h}\times\boldsymbol{F}+\boldsymbol{m}\right).
\end{align*}
These equations close if the constitutive laws and boundary conditions
only depend on the deformation. Projecting the above in the moving
frame and using the identities $\boldsymbol{e}_{i}\cdot\boldsymbol{A}^{\prime}=\left(\underline{A}^{\prime}+\text{\underbar{\ensuremath{\Pi}}\ensuremath{\times\underline{A}}}\right)_{i}$
and $\boldsymbol{e}_{i}\cdot\boldsymbol{\dot{A}}=\left(\underline{\dot{A}}+\text{\underbar{\ensuremath{\Omega}}\ensuremath{\times\underline{A}}}\right)_{i}$,
we obtain 
\begin{equation}
\dot{E}=\mathcal{N}\left(E\right),\quad E=\left(h_{1},h_{2},h_{3},\Pi_{1},\Pi_{2},\Pi_{3}\right),\label{eq:shape_space_dyn}
\end{equation}
where $\mathcal{N}$ is a reaction--advection--diffusion (RAD) operator
whose explicit form depends on the constitutive laws. This is a six-species
pattern forming system in $1+1$ dimensional spacetime and describes
the geometric flow of the shape field $E(u,t)$. Diffusion arises
from the elastic stresses while, crucially, geometric nonlinearities
generate advection and reaction terms. The free boundary conditions
depend only on $E$ and the system closes in shape space.

\textit{Mechanically linear media. }We now specify constitutive laws.
Following \citep{antman_nonlinear_2004}, we take the harmonic elastic
energy 
\begin{equation}
\mathcal{E}=\frac{1}{2}\int_{0}^{L}du\left[\left(\boldsymbol{h}-\boldsymbol{e}_{1}\right)^{\text{tr}}\boldsymbol{k}^{T}\left(\boldsymbol{h}-\boldsymbol{e}_{1}\right)+\boldsymbol{\Pi}^{\text{tr}}\boldsymbol{k}^{R}\boldsymbol{\Pi}\right],
\end{equation}
which yields linear stress constitutive laws $\boldsymbol{F}=\boldsymbol{k}^{T}\cdot\left(\boldsymbol{h}-\boldsymbol{e}_{1}\right)$,
$\boldsymbol{M}=\boldsymbol{k}^{R}\cdot\boldsymbol{\Pi}$. We assume
that the stiffness tensors simultaneously diagonalize in the moving
frame so that $\underline{\underline{k}}^{T}=\mathrm{diag}(k_{\parallel}^{T},k_{\perp}^{T},k_{\perp}^{T})$,
$\underline{\underline{k}}^{R}=\mathrm{diag}(k_{\parallel}^{R},k_{\perp}^{R},k_{\perp}^{R})$.
\begin{figure}
\centering
\includegraphics[width=1\columnwidth]{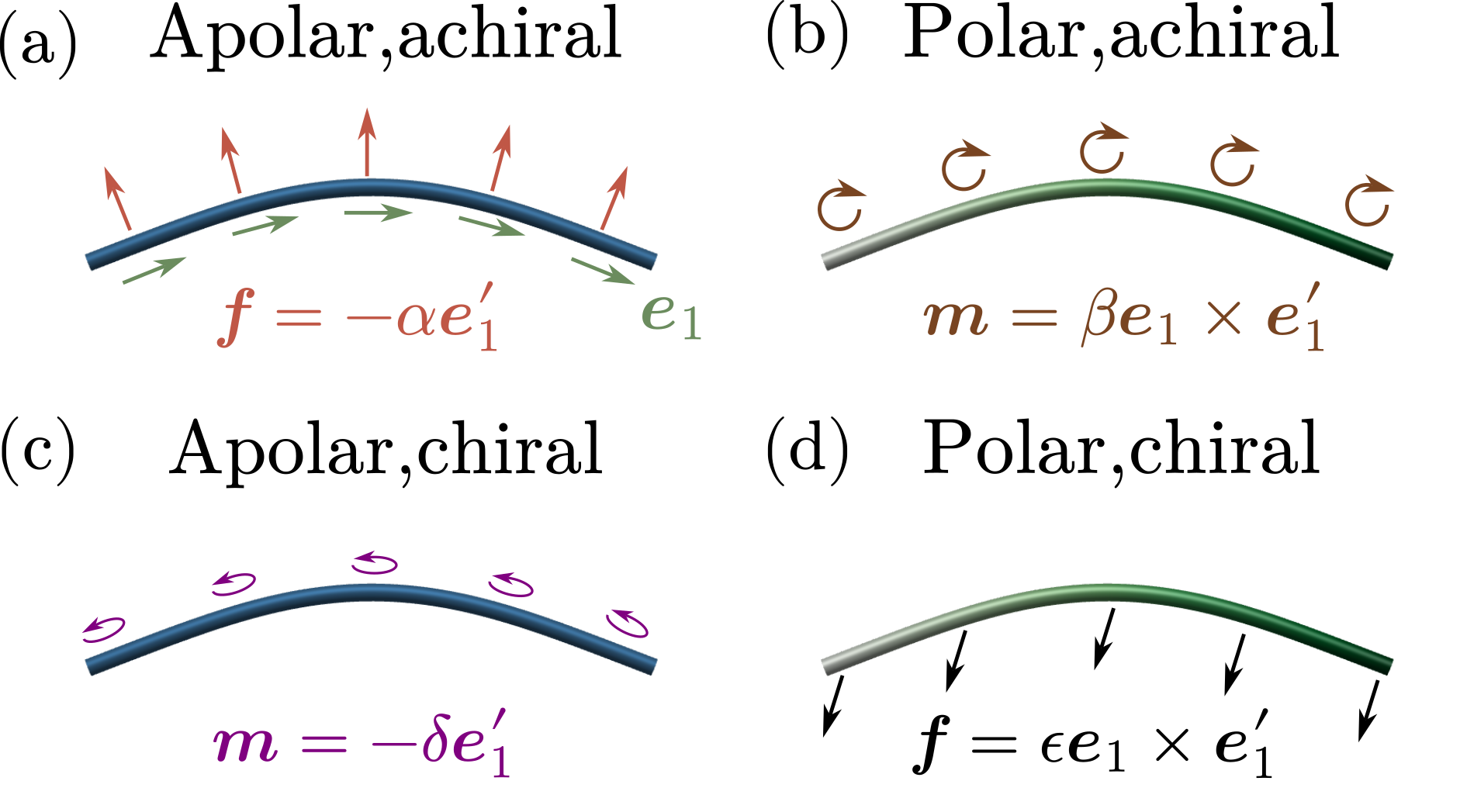} \caption{Active force and torque densities \eqref{eq:explicit_densities} and
their symmetry classification \citep{nemeth_nonreciprocal_2026}.
(a) Normal force density proportional to curvature, similar to force
densities in active nematics \citep{aditi_simha_hydrodynamic_2002}.
(b) Polar, achiral torque density proportional to curvature, leading
to bend propagation \citep{blum_biophysics_1979}. (c) Apolar, chiral
torque density, arising from, e.g., microscopic torque dipoles \citep{alexander_screw_2025}.
(d) Polar, chiral force density: an ``odd'' bending modulus generating
transverse response to bending deformation.}
\end{figure}

Active contributions $\boldsymbol{f},\boldsymbol{m}$ arise from nonreciprocal
interactions, e.g. coarse-grained hydrodynamic or chemical fields
\citep{jayaraman_autonomous_2012,butler_elastohydrodynamics_2026}.
The simplest strain-linear, transversely-isotropic form vanishing
for straight configurations is \citep{nemeth_nonreciprocal_2026}
\begin{equation}
\boldsymbol{f}=-\alpha\boldsymbol{e}_{1}'+\epsilon\boldsymbol{e}_{1}\times\boldsymbol{e}_{1}',\quad\boldsymbol{m}=-\delta\boldsymbol{e}_{1}'+\beta\boldsymbol{e}_{1}\times\boldsymbol{e}_{1}'.\label{eq:explicit_densities}
\end{equation}
Here $\beta,\epsilon$ break left--right symmetry, while $\delta,\epsilon$
break mirror symmetry in the plane of the cross-sections. Although
some terms are total derivatives, they are not equivalent to prestress
due to boundary conditions. The RAD system corresponding to these
constitutive choices is derived in the SM \citep{sm}. We now study
its fixed points and bifurcations. 
\begin{figure*}
\centering
\includegraphics[width=1\textwidth]{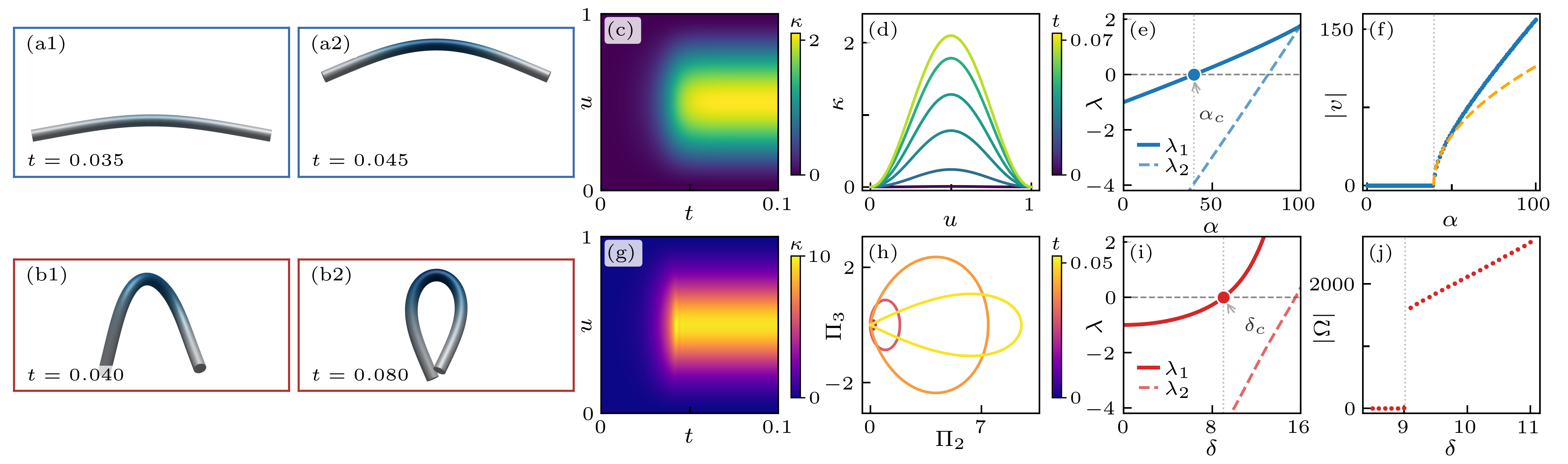} \caption{Fixed points of shape dynamics (see Movie 1 of the SM \citep{sm}
for animations of the fixed point behavior). (a1)--(a2): $\alpha=50$,
self-propelling U-shape. The rod is shaded according to its scalar curvature. (b1)--(b2): $\delta=10$, rotating hairpin
shape. (c): Curvature kymograph of a rod with $\alpha=50$. (d) Evolution of the
curvature profile of the rod with $\alpha=50$. (e) Leading eigenvalues
of Eq. \eqref{eq:linearization} as a function of $\alpha$, with
other parameters zero. (f) Self-propulsion velocity from simulations
as a function of $\alpha$. The dashed line is the prediction $v\approx14.7\sqrt{\alpha-4\pi^{2}}$
from a weakly nonlinear analysis. (g) Curvature kymograph of a rod
with $\delta=10$. (h) Evolution of the shape of the rod with $\delta=10$.
(i) Leading eigenvalues of Eq.~\eqref{eq:linearization} as a function
of $\delta$, with other parameters zero. (j) Angular velocity from
simulations as a function of $\delta$.}
\end{figure*}

\textit{Fixed points (stationary attractors).} In the absence of nonreciprocity,
the homogeneous shape $\underline{h}=\text{\ensuremath{\underline{e_{1}}}},\text{\underbar{\ensuremath{\Pi}}}=\underline{0}$,
corresponding to the straight configuration, is the unique fixed point
of Eq.~\eqref{eq:shape_space_dyn} and is linearly stable. Within
the shape-space formulation, this state is a homogeneous stationary
attractor. Upon reconstruction, such attractors correspond to rigid
configurations with no intrinsic deformation or self-propulsion.

Nonreciprocal forcing alters this structure by generating curvature
through the $\alpha$ and $\delta$ terms. As these coefficients are
increased, the homogeneous attractor loses stability, giving rise
to new inhomogeneous steady states in shape space, see Fig.~2. These
emergent attractors correspond, upon reconstruction, to time-independent
deformed configurations undergoing rigid motion in physical space.

To analyze the onset of instabilities, we linearize the shape equation
about the homogeneous attractor. We nondimensionalize by rescaling
lengths by $L$ and times by the bending timescale $T=\gamma^{T}L^{4}/k_{\perp}^{R}$.
In the limit of large stretch, shear and twist moduli, $k_{\perp}^{R}\ll k_{\parallel}^{R},k_{\parallel}^{T}L^{2},k_{\perp}^{T}L^{2}$
and $\gamma^{R}\ll\gamma^{T}L^{2}$, the dynamics of $\underline{h}$
and $\Pi_{1}$ separate in time scale from the dynamics of $\Pi_{2}$
and $\Pi_{3}$. This defines a slow manifold corresponding to the
filament limit, in which the rod is inextensible ($|\boldsymbol{r}'|=1$)
and unshearable ($\boldsymbol{r}'\parallel\boldsymbol{e}_{1}$). In
this regime, the dynamics is naturally expressed in terms of the complex
curvature $\Pi=\Pi_{2}+i\Pi_{3}$, which encodes both bending and
torsion of the centerline. 

On this slow manifold, stability is governed by the linear operator
\begin{equation}
\dot{\Pi}=-\Pi''''-\mu\Pi'''-\nu\Pi'',\quad\Pi=\Pi_{2}+i\Pi_{3},\label{eq:linearization}
\end{equation}
where $\mu=\left(\beta+i\delta\right)\times L/k_{\perp}^{R}$ and
$\nu=\left(\alpha-i\epsilon\right)\times L^{2}/k_{\perp}^{R}$ are
nondimensional complex-valued parameters encoding the strength of
nonreciprocal forcing. The stability of the homogeneous attractor
is determined by the spectrum of this operator with boundary conditions
$\Pi=\Pi'=0$ at both endpoints, which predicts the onset of instabilities
at critical values $\alpha_{c}=4\pi^{2}$ and $\delta_{c}\approx8.99$,
in excellent agreement with numerical simulations.

Beyond these thresholds, the stationary attractor loses stability
and is replaced by new attractors in shape space. For $\alpha>\alpha_{c}$,
a supercritical pitchfork bifurcation produces a branch of self-propelling
$U$-shaped configurations, consistently with recent theoretical and
experimental observations in active colloidal chains and chemically
active filaments \citep{jayaraman_autonomous_2012,prathyusha_emergent_2022,lough_self-buckling_2023,kumar_emergent_2024,makanga_instability_2026}.
Weakly nonlinear analysis shows that the associated translational
velocity scales as $v\simeq14.7\sqrt{\alpha-\alpha_{c}}$ near onset,
see Fig.~2(f). Averaging the equations of motion over a planar shape
self-propelling with speed $v$, we find the relation 
\begin{equation}
v=2\alpha\sin\left(\bar{\kappa}/2\right)\label{eq:prop_curv}
\end{equation}
between the average scalar curvature $\bar{\kappa}=\int_{0}^{1}\kappa du$
of the rod and its speed, yielding an exact nonlinear generalization
of the result reported in \citep{jayaraman_autonomous_2012}.

For an apolar, chiral rod with $\delta\neq0,\alpha=\beta=\epsilon=0$,
a distinct instability leads to rotating hairpin structures associated
with global rigid rotation in physical space. With increasing $\delta$,
we observe a helical buckling instability at a critical value $\delta_{c}\approx9$
\citep{landau_theory_1984,goldstein_viscous_1998,nemeth_nonreciprocal_2026},
after which the rod settles into a rotating hairpin that undergoes
global rigid rotation, with the amplitude jumping discontinuously
at the bifurcation. In both cases, the new steady states remain fixed
points in shape space but correspond to nontrivial rigid motion upon
reconstruction.
\begin{figure*}
\centering
\includegraphics[width=1\textwidth]{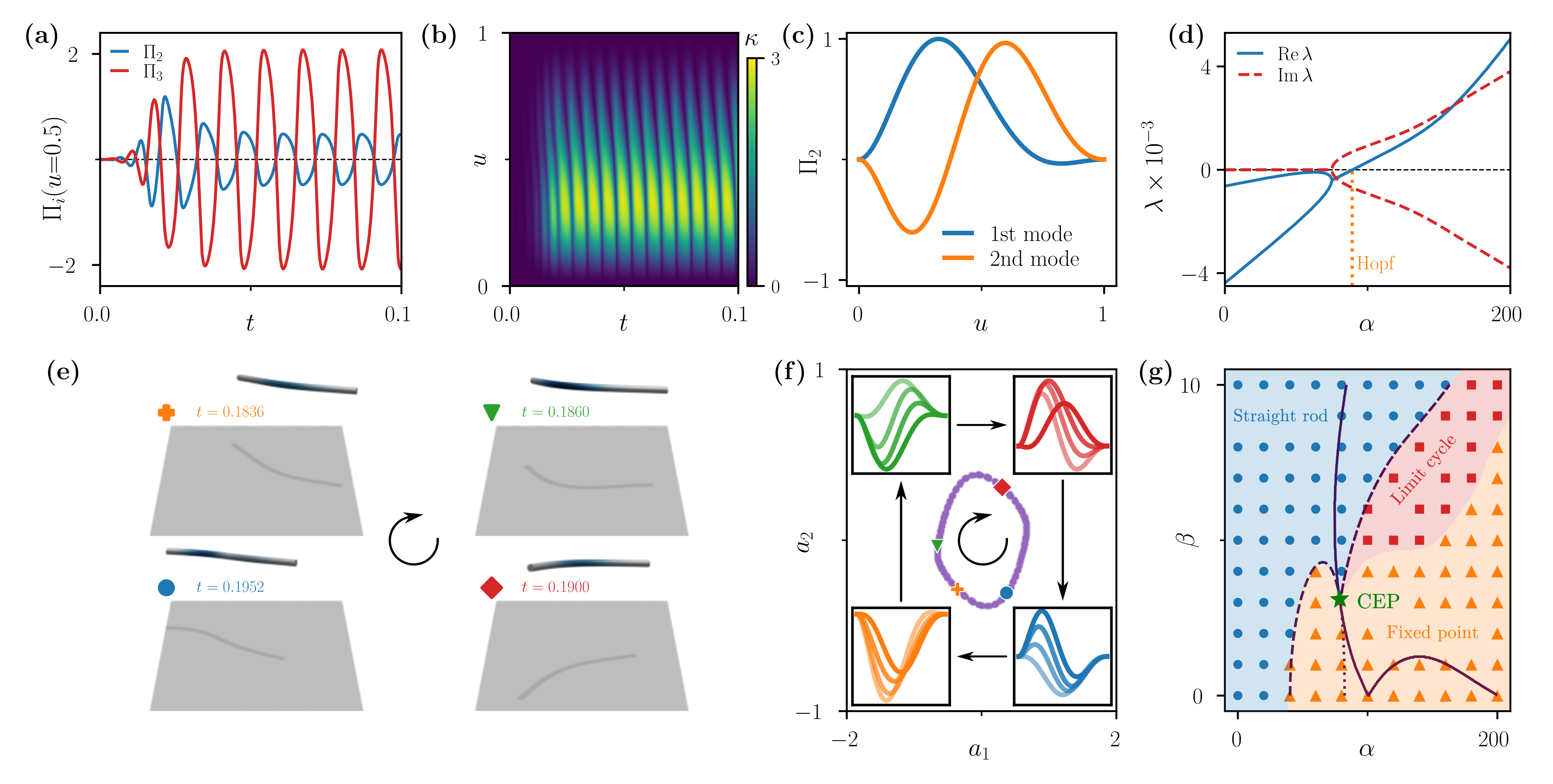}

\caption{Curvature evolution of a polar, achiral rod with $\alpha=100,\beta=5$, exhibiting a limit cycle
in shape space. (a) Temporal evolution of $\Pi_{2},\Pi_{3}$ the midpoint
of the rod. (b) Curvature kymograph of the rod. (c) Dominant $\Pi_{2}$ modes obtained
from principal component analysis (PCA). (d) Real and imaginary parts
of the leading two eigenvalues of the dynamics at $\beta=5$ as a
function of $\alpha$. (e) Snapshots of the rod dynamics over a limit
cycle. The rod is shaded according to its scalar curvature. The gray
plane is a guide to the eye and does not represent a physical boundary.
See Movie 2 of the SM \citep{sm} for animations of the limit cycle.
(f) $\Pi_{2}$ dynamics of the rod projected onto the subspace spanned
by the two dominant PCA modes, exhibiting a limit cycle. The insets
show the evolution of $\Pi_{2}$ along the limit cycle. (g) Phase
diagram of a polar, achiral rod. The dashed and dotted lines are the
curves where the first and the second leading eigenvalues of the linearized
dynamics cross the real axis, while the solid line is where they acquire
an imaginary part. The intersection near $\alpha\approx77,\beta\approx3$
of the three curves is a critical exceptional point. Crossing from the blue to
the red region of the phase diagram upon increasing $\alpha$, we
find a characteristic Hopf bifurcation. Reentrant behavior is observed
for $\beta\approx4$.}
\end{figure*}

The fixed points in shape space are robust to perturbations in the
parameters but acquire qualitatively new features depending on the
symmetry of the rod. At a fixed $\alpha>\alpha_{c},\delta=\epsilon=0$,
breaking polar symmetry by a small $\beta$ tilts the stationary $U$-shape,
leading to a net torque on the rod and planar global rotating motion,
while a small $\epsilon$ at $\alpha>\alpha_{c},\beta=\delta=0$ adds
transverse forces, giving rise to chiral out-of-plane rotating motion.
Higher modes also become unstable when $\alpha$ is increased further,
and we find rotating stationary shapes even in the absence of polar
or chiral terms \citep{sm}.

\textit{Limit cycles (periodic attractors).} Beyond stationary attractors,
RAD systems generically support oscillatory attractors arising through
Hopf bifurcations. Exploring the nonreciprocal parameter space, we
identify a regime in which the shape dynamics of a polar, achiral
rod ($\alpha,\beta\neq0$, $\delta=\epsilon=0$) evolves onto a stable
limit cycle, see Fig.~3.

As $\alpha$ is increased at fixed $\beta\apprge5$, a pair of complex-conjugate
eigenvalues of the linearized dynamics crosses the imaginary axis,
see Fig.~3(d), and the stationary attractor loses stability. The
system is thereby replaced by a periodic attractor in shape space,
corresponding to a time-periodic deformation pattern involving coordinated
oscillations of curvature modes along the filament, see Fig.~3(a)-(b).

The origin of this transition can be understood from the interplay
of curvature production and transport. While the $\alpha$ term promotes
curvature generation, the polar coefficient $\beta$ introduces advective
coupling in the shape dynamics, producing an imaginary component in
the spectrum of the linearized operator. Together with elastic relaxation,
this provides a generic mechanism for oscillatory instability. Fig.~3(g)
summarizes the resulting phase diagram. The Hopf bifurcation governs
the onset of oscillations, while the intersection of instability boundaries
near $\alpha\approx77,\beta\approx3$ corresponds to a critical exceptional
point \citep{fruchart_non-reciprocal_2021,al-izzi_nonreciprocal_2026}
where two eigenvalues simultaneously go to zero and the eigenspace
becomes degenerate, with the three phases (straight rod, nontrivial fixed point, limit cycle) meeting there.

\begin{figure*}
\centering
\includegraphics[width=1\textwidth]{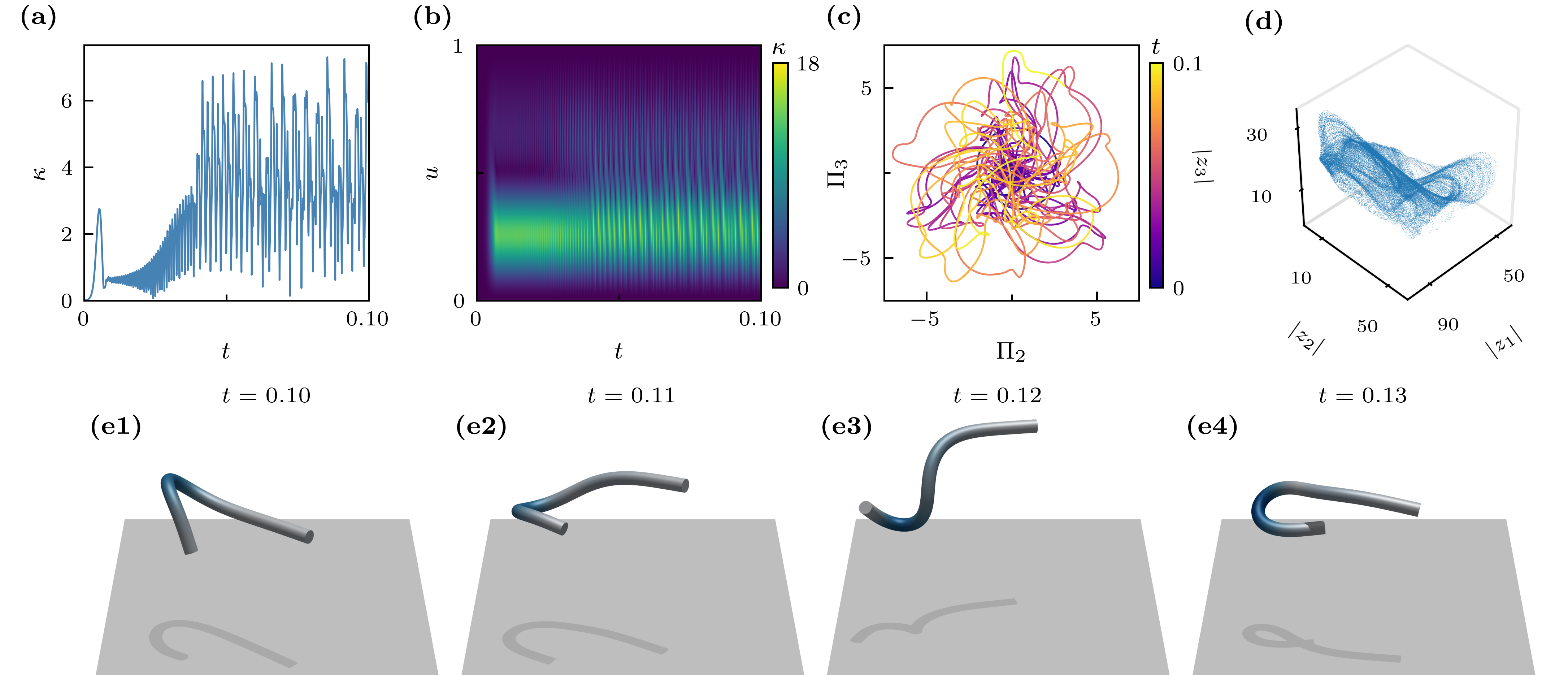}\caption{Chaotic dynamics for $\delta=15,\beta=3.5$. (a) Temporal evolution
of the curvature of the midpoint of the rod. (b) Curvature kymograph
of the rod. (c) Evolution of the shape of the rod at the midpoint
in $\Pi_{2}-\Pi_{3}$ space. (d) Chaotic attractor formed from scatter
plot of dynamics projected onto the amplitudes of the first three
dominant eigenmodes obtained from PCA. (e1)--(e4) Snapshots of chaotic
rod dynamics. See Movie 3 of the SM \citep{sm} for animations of
chaotic dynamics.}
\end{figure*}
Upon reconstruction, periodic attractors in shape space correspond
to autonomous swimming gaits in physical space, see Fig.~3(e). Each
cycle in shape space produces a net displacement through the kinematic
coupling between deformation and rigid motion, in analogy with the
geometric phase underlying low-Reynolds-number locomotion \citep{shapere_geometry_1989,kanso_swimming_2009,rieser_geometric_2024}.
Unlike classical geometric descriptions \citep{montgomery_gauge_1993},
however, the limit cycle is not prescribed but emerges dynamically
through a bifurcation due to the nonreciprocity of the medium. The
resulting motion is therefore an autonomous gait selected by the intrinsic
shape dynamics. We note that left--right symmetry breaking is not
essential: oscillatory attractors persist even in apolar, achiral
rods at sufficiently large activity \citep{sm}. 

\textit{Chaos (chaotic attractors).} Periodic attractors of RAD systems
generically lose stability to chaotic attractors arising from the
interaction of instability, transport, and dissipation. Further exploring
the nonreciprocal parameter space, we identify a regime in which the
shape dynamics of a polar, chiral rod ($\beta,\delta\neq0,\alpha=\epsilon=0$)
evolves onto a chaotic attractor, see Fig.~4. In this regime, the
curvature field exhibits seemingly chaotic spatiotemporal structure,
while projections onto the dominant principal components reveal an
effective low-dimensional bounded chaotic trajectory in shape space.

The onset of chaos arises from the competing roles of the nonreciprocal
contributions. The chiral coefficient $\delta$ destabilizes the straight
configuration and continuously injects curvature into the filament,
while the polar coefficient $\beta$ advects curvature fluctuations
along its length. In a finite filament, these disturbances are repeatedly
reflected at the free boundaries and interact nonlinearly with newly
generated curvature waves. Upon reconstruction, chaotic attractors
in shape space correspond to irregular trajectories in physical space,
characterized by aperiodic sequences of bending, rotation, and writhing
driven by the underlying deformation dynamics. The resulting motion
is therefore neither steady nor periodic, but governed by chaotic
evolution in shape space \citep{krishnamurthy_emergent_2023,sarkar_mechanochemical_2025,bonacci_reversibility_2026}.

This mechanism closely parallels other pattern-forming systems exhibiting
deterministic chaos, such as the Kuramoto--Sivashinsky equation \citep{kuramoto_persistent_1976,sivashinsky_nonlinear_1977},
where unstable long-wavelength modes are balanced by higher-order
dissipation. Here, however, the chaotic dynamics unfolds in shape
space and is mapped to physical motion through the kinematics.

\textit{Discussion.} We have shown that the elastohydrodynamics of
nonreciprocal slender solids admits a closed description in shape
space, where Euclidean symmetry is factored out and deformation evolves
as a nonlinear RAD system. This reformulates nonreciprocal active
elastohydrodynamics as a pattern-formation problem: fixed points,
limit cycles, and chaotic attractors of the shape dynamics manifest
as rigid motion, autonomous swimming gaits, and chaotic locomotion
in physical space. In contrast to classical geometric theories of
swimming, where shape changes are prescribed, the relevant trajectories
in shape space emerge spontaneously through instabilities of the nonreciprocally
active medium.

Although we focused on the simplest local nonreciprocal force and
torque densities compatible with Euclidean invariance, the bifurcations
reported here appear robust and show qualitative agreement with detailed
hydrodynamic models of nonreciprocal active slender bodies \citep{jayaraman_autonomous_2012,butler_elastohydrodynamics_2026,makanga_instability_2026}.
This suggests that the observed phenomenology reflects generic dynamical
structures of nonreciprocal slender bodies rather than specific constitutive
choices. The appearance of Hopf bifurcations, chaos, and exceptional
points further points to connections with pattern formation and non-Hermitian
dynamics \citep{ashida_non-hermitian_2020}. Bifurcations in shape
space dynamics can be harnessed via spatiotemporal modulation of the
nonreciprocal coefficients to control the shape and motion of nonreciprocal
filaments \citep{qiao_control_2022}.

The shape-space formulation also opens several analytical directions.
Nonlocal hydrodynamic interactions and fluctuations can be incorporated
systematically, while the reaction--advection--diffusion structure
provides a natural starting point for amplitude-equation and normal-form
descriptions near instability thresholds \citep{cross_pattern_1993}.
It may also offer a route toward connecting nonreciprocal filament
dynamics with geometric evolution equations and classical theories
of curve dynamics \citep{hasimoto_soliton_1972,cohen_schrodinger_2023}.

\textit{Acknowledgements.} We thank Anton Souslov, Brato Chakrabarti
and Rajesh Singh for stimulating discussions and helpful suggestions.
This research was supported by Engineering and Physical Sciences Research
Council Grant No. EP/W524141/1 (B.N.) and W108115D (M.W.) B.N. and
M.W. contributed equally to this work.

\textit{Data availability.} The data supporting the findings of this
study are available within the article.

\nocite{*}
\bibliography{manuscript}

@article{poncet_when_2022,
	title = {When {Soft} {Crystals} {Defy} {Newton}’s {Third} {Law}: {Nonreciprocal} {Mechanics} and {Dislocation} {Motility}},
	volume = {128},
	issn = {0031-9007, 1079-7114},
	shorttitle = {When {Soft} {Crystals} {Defy} {Newton}’s {Third} {Law}},
	url = {https://link.aps.org/doi/10.1103/PhysRevLett.128.048002},
	doi = {10.1103/PhysRevLett.128.048002},
	number = {4},
	urldate = {2025-05-20},
	journal = {Physical Review Letters},
	author = {Poncet, Alexis and Bartolo, Denis},
	month = jan,
	year = {2022},
	pages = {048002},
}

@book{landau_theory_1984,
	address = {Oxford, England},
	edition = {3},
	title = {Theory of elasticity},
	publisher = {Butterworth-Heinemann},
	author = {Landau, L. D. and Pitaevskii, L. P. and Lifshitz, E. M. and Kosevich, A. M.},
	year = {1984},
}

@article{scheibner_odd_2020,
	title = {Odd elasticity},
	volume = {16},
	issn = {1745-2473, 1745-2481},
	url = {https://www.nature.com/articles/s41567-020-0795-y},
	doi = {10.1038/s41567-020-0795-y},
	number = {4},
	urldate = {2025-06-28},
	journal = {Nature Physics},
	author = {Scheibner, Colin and Souslov, Anton and Banerjee, Debarghya and Surówka, Piotr and Irvine, William T. M. and Vitelli, Vincenzo},
	month = apr,
	year = {2020},
	pages = {475--480},
}

@article{wiggins_flexive_1998,
	title = {Flexive and {Propulsive} {Dynamics} of {Elastica} at {Low} {Reynolds} {Number}},
	volume = {80},
	copyright = {http://link.aps.org/licenses/aps-default-license},
	issn = {0031-9007, 1079-7114},
	url = {https://link.aps.org/doi/10.1103/PhysRevLett.80.3879},
	doi = {10.1103/PhysRevLett.80.3879},
	number = {17},
	urldate = {2025-09-03},
	journal = {Physical Review Letters},
	author = {Wiggins, Chris H. and Goldstein, Raymond E.},
	month = apr,
	year = {1998},
	pages = {3879--3882},
}

@book{antman_nonlinear_2004,
	address = {New York},
	title = {Nonlinear {Problems} of {Elasticity}},
	publisher = {Springer},
	author = {Antman, S.S.},
	year = {2004},
}

@article{jayaraman_autonomous_2012,
	title = {Autonomous {Motility} of {Active} {Filaments} due to {Spontaneous} {Flow}-{Symmetry} {Breaking}},
	volume = {109},
	copyright = {http://link.aps.org/licenses/aps-default-license},
	issn = {0031-9007, 1079-7114},
	url = {https://link.aps.org/doi/10.1103/PhysRevLett.109.158302},
	doi = {10.1103/PhysRevLett.109.158302},
	number = {15},
	urldate = {2025-10-20},
	journal = {Physical Review Letters},
	author = {Jayaraman, Gayathri and Ramachandran, Sanoop and Ghose, Somdeb and Laskar, Abhrajit and Bhamla, M. Saad and Kumar, P. B. Sunil and Adhikari, R.},
	month = oct,
	year = {2012},
	pages = {158302},
}

@article{goldstein_nonlinear_1995,
	title = {Nonlinear {Dynamics} of {Stiff} {Polymers}},
	volume = {75},
	copyright = {http://link.aps.org/licenses/aps-default-license},
	issn = {0031-9007, 1079-7114},
	url = {https://link.aps.org/doi/10.1103/PhysRevLett.75.1094},
	doi = {10.1103/PhysRevLett.75.1094},
	number = {6},
	urldate = {2025-10-20},
	journal = {Physical Review Letters},
	author = {Goldstein, Raymond E. and Langer, Stephen A.},
	month = aug,
	year = {1995},
	pages = {1094--1097},
}

@article{shapere_geometry_1989,
	title = {Geometry of self-propulsion at low {Reynolds} number},
	volume = {198},
	copyright = {https://www.cambridge.org/core/terms},
	issn = {0022-1120, 1469-7645},
	url = {https://www.cambridge.org/core/product/identifier/S002211208900025X/type/journal_article},
	doi = {10.1017/S002211208900025X},
	urldate = {2025-10-20},
	journal = {Journal of Fluid Mechanics},
	author = {Shapere, Alfred and Wilczek, Frank},
	month = jan,
	year = {1989},
	pages = {557--585},
}

@article{audoly_fragmentation_2005,
	title = {Fragmentation of {Rods} by {Cascading} {Cracks}: {Why} {Spaghetti} {Does} {Not} {Break} in {Half}},
	volume = {95},
	copyright = {http://link.aps.org/licenses/aps-default-license},
	issn = {0031-9007, 1079-7114},
	shorttitle = {Fragmentation of {Rods} by {Cascading} {Cracks}},
	url = {https://link.aps.org/doi/10.1103/PhysRevLett.95.095505},
	doi = {10.1103/PhysRevLett.95.095505},
	number = {9},
	urldate = {2025-12-11},
	journal = {Physical Review Letters},
	author = {Audoly, Basile and Neukirch, Sébastien},
	month = aug,
	year = {2005},
	pages = {095505},
}

@book{audoly_elasticity_2010,
	title = {Elasticity and geometry: from hair curls to the non-linear response of shells},
	isbn = {978-0-19-850625-6 978-0-19-154502-3},
	shorttitle = {Elasticity and geometry},
	publisher = {Oxford University Press},
	editor = {Audoly, Basile and Pomeau, Yves},
	year = {2010},
}

@book{euler_methodus_1744,
	address = {Lausanne \& Geneva},
	title = {Methodus inveniendi lineas curvas maximi minive proprietate gaudentes},
	publisher = {Bousquet},
	author = {Euler, Leonhard},
	year = {1744},
}

@article{lough_self-buckling_2023,
	title = {Self-buckling and self-writhing of semi-flexible microorganisms},
	volume = {19},
	issn = {1744-683X, 1744-6848},
	url = {https://xlink.rsc.org/?DOI=D3SM00572K},
	doi = {10.1039/D3SM00572K},
	number = {38},
	urldate = {2025-12-11},
	journal = {Soft Matter},
	author = {Lough, Wilson and Weibel, Douglas B. and Spagnolie, Saverio E.},
	year = {2023},
	pages = {7349--7357},
}

@article{prathyusha_emergent_2022,
	title = {Emergent conformational properties of end-tailored transversely propelling polymers},
	volume = {18},
	issn = {1744-683X, 1744-6848},
	url = {https://xlink.rsc.org/?DOI=D2SM00237J},
	doi = {10.1039/D2SM00237J},
	number = {15},
	urldate = {2025-12-11},
	journal = {Soft Matter},
	author = {Prathyusha, K. R. and Ziebert, Falko and Golestanian, Ramin},
	year = {2022},
	pages = {2928--2935},
}

@article{ishimoto_self-organized_2022,
	title = {Self-organized swimming with odd elasticity},
	volume = {105},
	issn = {2470-0045, 2470-0053},
	url = {https://link.aps.org/doi/10.1103/PhysRevE.105.064603},
	doi = {10.1103/PhysRevE.105.064603},
	number = {6},
	urldate = {2025-12-11},
	journal = {Physical Review E},
	author = {Ishimoto, Kenta and Moreau, Clément and Yasuda, Kento},
	month = jun,
	year = {2022},
	pages = {064603},
}

@article{winkler_physics_2020,
	title = {The physics of active polymers and filaments},
	volume = {153},
	issn = {0021-9606, 1089-7690},
	url = {https://pubs.aip.org/jcp/article/153/4/040901/957885/The-physics-of-active-polymers-and-filaments},
	doi = {10.1063/5.0011466},
	number = {4},
	urldate = {2025-12-11},
	journal = {The Journal of Chemical Physics},
	author = {Winkler, Roland G. and Gompper, Gerhard},
	month = jul,
	year = {2020},
	pages = {040901},
}

@article{moulton_multiscale_2020,
	title = {Multiscale integration of environmental stimuli in plant tropism produces complex behaviors},
	volume = {117},
	issn = {0027-8424, 1091-6490},
	url = {https://pnas.org/doi/full/10.1073/pnas.2016025117},
	doi = {10.1073/pnas.2016025117},
	number = {51},
	urldate = {2025-12-11},
	journal = {Proceedings of the National Academy of Sciences},
	author = {Moulton, Derek E. and Oliveri, Hadrien and Goriely, Alain},
	month = dec,
	year = {2020},
	pages = {32226--32237},
}

@article{timoshenko_lxvi_1921,
	title = {{LXVI}. {On} the correction for shear of the differential equation for transverse vibrations of prismatic bars},
	volume = {41},
	issn = {1941-5982, 1941-5990},
	url = {https://www.tandfonline.com/doi/full/10.1080/14786442108636264},
	doi = {10.1080/14786442108636264},
	number = {245},
	urldate = {2025-12-11},
	journal = {The London, Edinburgh, and Dublin Philosophical Magazine and Journal of Science},
	author = {Timoshenko, S.P.},
	month = may,
	year = {1921},
	pages = {744--746},
}

@article{kirchhoff_uber_1859,
	title = {Über das {Gleichgewicht} und die {Bewegung} eines unendlich dünnen elastischen {Stabes}},
	volume = {56},
	journal = {Journal für die reine und angewandte Mathematik},
	author = {Kirchhoff, Gustav},
	year = {1859},
	pages = {285--313},
}

@article{goldstein_viscous_1998,
	title = {Viscous {Nonlinear} {Dynamics} of {Twist} and {Writhe}},
	volume = {80},
	copyright = {http://link.aps.org/licenses/aps-default-license},
	issn = {0031-9007, 1079-7114},
	url = {https://link.aps.org/doi/10.1103/PhysRevLett.80.5232},
	doi = {10.1103/PhysRevLett.80.5232},
	number = {23},
	urldate = {2026-05-25},
	journal = {Physical Review Letters},
	author = {Goldstein, Raymond E. and Powers, Thomas R. and Wiggins, Chris H.},
	month = jun,
	year = {1998},
	pages = {5232--5235},
}

@article{nemeth_nonreciprocal_2026,
	title = {Nonreciprocal constitutive laws for oriented active solids},
	volume = {28},
	issn = {1367-2630},
	url = {https://iopscience.iop.org/article/10.1088/1367-2630/ae4f19},
	doi = {10.1088/1367-2630/ae4f19},
	number = {3},
	urldate = {2026-05-25},
	journal = {New Journal of Physics},
	author = {Németh, Balázs and Kobayashi, Takuya and Adhikari, Ronojoy},
	month = mar,
	year = {2026},
	pages = {034401},
}

@article{wei_life-like_2026,
	title = {Life-like behavior emerging in active and flexible microstructures},
	volume = {123},
	issn = {0027-8424, 1091-6490},
	url = {https://pnas.org/doi/10.1073/pnas.2531743123},
	doi = {10.1073/pnas.2531743123},
	number = {13},
	urldate = {2026-05-25},
	journal = {Proceedings of the National Academy of Sciences},
	author = {Wei, Mengshi and Kraft, Daniela J.},
	month = mar,
	year = {2026},
	pages = {e2531743123},
}

@article{cass_reaction-diffusion_2023,
	title = {The reaction-diffusion basis of animated patterns in eukaryotic flagella},
	volume = {14},
	issn = {2041-1723},
	url = {https://www.nature.com/articles/s41467-023-40338-2},
	doi = {10.1038/s41467-023-40338-2},
	number = {1},
	urldate = {2026-05-25},
	journal = {Nature Communications},
	author = {Cass, James F. and Bloomfield-Gadêlha, Hermes},
	month = sep,
	year = {2023},
	pages = {5638},
}

@article{kumar_emergent_2024,
	title = {Emergent dynamics due to chemo-hydrodynamic self-interactions in active polymers},
	volume = {15},
	issn = {2041-1723},
	url = {https://www.nature.com/articles/s41467-024-49155-7},
	doi = {10.1038/s41467-024-49155-7},
	number = {1},
	urldate = {2026-05-25},
	journal = {Nature Communications},
	author = {Kumar, Manoj and Murali, Aniruddh and Subramaniam, Arvin Gopal and Singh, Rajesh and Thutupalli, Shashi},
	month = jun,
	year = {2024},
	pages = {4903},
}

@article{cross_pattern_1993,
	title = {Pattern formation outside of equilibrium},
	volume = {65},
	copyright = {http://link.aps.org/licenses/aps-default-license},
	issn = {0034-6861, 1539-0756},
	url = {https://link.aps.org/doi/10.1103/RevModPhys.65.851},
	doi = {10.1103/RevModPhys.65.851},
	number = {3},
	urldate = {2026-05-25},
	journal = {Reviews of Modern Physics},
	author = {Cross, M. C. and Hohenberg, P. C.},
	month = jul,
	year = {1993},
	pages = {851--1112},
}

@article{veenstra_adaptive_2025,
	title = {Adaptive locomotion of active solids},
	volume = {639},
	issn = {0028-0836, 1476-4687},
	url = {https://www.nature.com/articles/s41586-025-08646-3},
	doi = {10.1038/s41586-025-08646-3},
	number = {8056},
	urldate = {2026-05-25},
	journal = {Nature},
	author = {Veenstra, Jonas and Scheibner, Colin and Brandenbourger, Martin and Binysh, Jack and Souslov, Anton and Vitelli, Vincenzo and Coulais, Corentin},
	month = mar,
	year = {2025},
	pages = {935--941},
}

@article{al-izzi_nonreciprocal_2026,
	title = {Nonreciprocal buckling makes active filaments polyfunctional},
	volume = {123},
	issn = {0027-8424, 1091-6490},
	url = {https://pnas.org/doi/10.1073/pnas.2531723123},
	doi = {10.1073/pnas.2531723123},
	number = {11},
	urldate = {2026-05-25},
	journal = {Proceedings of the National Academy of Sciences},
	author = {Al-Izzi, Sami C. and Du, Yao and Veenstra, Jonas and Morris, Richard G. and Souslov, Anton and Carlson, Andreas and Coulais, Corentin and Binysh, Jack},
	month = mar,
	year = {2026},
	pages = {e2531723123},
}

@article{chen_realization_2021,
	title = {Realization of active metamaterials with odd micropolar elasticity},
	volume = {12},
	issn = {2041-1723},
	url = {https://www.nature.com/articles/s41467-021-26034-z},
	doi = {10.1038/s41467-021-26034-z},
	number = {1},
	urldate = {2026-05-25},
	journal = {Nature Communications},
	author = {Chen, Yangyang and Li, Xiaopeng and Scheibner, Colin and Vitelli, Vincenzo and Huang, Guoliang},
	month = oct,
	year = {2021},
	pages = {5935},
}

@article{garg_slender_2023,
	title = {A slender body theory for the motion of special {Cosserat} filaments in {Stokes} flow},
	volume = {28},
	issn = {1081-2865, 1741-3028},
	url = {https://journals.sagepub.com/doi/10.1177/10812865221083323},
	doi = {10.1177/10812865221083323},
	number = {3},
	urldate = {2026-05-25},
	journal = {Mathematics and Mechanics of Solids},
	author = {Garg, Mohit and Kumar, Ajeet},
	month = mar,
	year = {2023},
	pages = {692--729},
}

@misc{bonacci_reversibility_2026,
	title = {Reversibility, {Chaos}, and {Attractors} in {Periodically} {Sheared} {Elastic} {Filaments}},
	url = {http://arxiv.org/abs/2601.00643},
	doi = {10.48550/arXiv.2601.00643},
	urldate = {2026-05-25},
	publisher = {arXiv},
	author = {Bonacci, Francesco and Chakrabarti, Brato and Roure, Olivia du and Lindner, Anke and Saintillan, David},
	month = jan,
	year = {2026},
	note = {arXiv:2601.00643 [cond-mat.soft]},
}

@article{chakrabarti_spontaneous_2019,
	title = {Spontaneous oscillations, beating patterns, and hydrodynamics of active microfilaments},
	volume = {4},
	issn = {2469-990X},
	url = {https://link.aps.org/doi/10.1103/PhysRevFluids.4.043102},
	doi = {10.1103/PhysRevFluids.4.043102},
	number = {4},
	urldate = {2026-05-25},
	journal = {Physical Review Fluids},
	author = {Chakrabarti, Brato and Saintillan, David},
	month = apr,
	year = {2019},
	pages = {043102},
}

@article{ericksen_exact_1957,
	title = {Exact theory of stress and strain in rods and shells},
	volume = {1},
	copyright = {http://www.springer.com/tdm},
	issn = {0003-9527, 1432-0673},
	url = {http://link.springer.com/10.1007/BF00298012},
	doi = {10.1007/BF00298012},
	number = {1},
	urldate = {2026-05-25},
	journal = {Archive for Rational Mechanics and Analysis},
	author = {Ericksen, J. L. and Truesdell, C.},
	month = jan,
	year = {1957},
	pages = {295--323},
}

@article{rieser_geometric_2024,
	title = {Geometric phase predicts locomotion performance in undulating living systems across scales},
	volume = {121},
	issn = {0027-8424, 1091-6490},
	url = {https://pnas.org/doi/10.1073/pnas.2320517121},
	doi = {10.1073/pnas.2320517121},
	number = {24},
	urldate = {2026-05-25},
	journal = {Proceedings of the National Academy of Sciences},
	author = {Rieser, Jennifer M. and Chong, Baxi and Gong, Chaohui and Astley, Henry C. and Schiebel, Perrin E. and Diaz, Kelimar and Pierce, Christopher J. and Lu, Hang and Hatton, Ross L. and Choset, Howie and Goldman, Daniel I.},
	month = jun,
	year = {2024},
	pages = {e2320517121},
}

@article{gelvan_hydrodynamic_2025,
	title = {Hydrodynamic spin-pairing and active polymerization of oppositely spinning rotors},
	volume = {16},
	issn = {2041-1723},
	url = {https://www.nature.com/articles/s41467-025-65322-w},
	doi = {10.1038/s41467-025-65322-w},
	number = {1},
	urldate = {2026-05-25},
	journal = {Nature Communications},
	author = {Gelvan, Mattan and Chirko, Artyom and Kirpitch, Jonathan and Lavie, Yahav and Israel, Noa and Oppenheimer, Naomi},
	month = nov,
	year = {2025},
	pages = {10368},
}

@article{chao_traveling_2025,
	title = {Traveling {Strings} of {Active} {Dipolar} {Colloids}},
	volume = {134},
	issn = {0031-9007, 1079-7114},
	url = {https://link.aps.org/doi/10.1103/PhysRevLett.134.018302},
	doi = {10.1103/PhysRevLett.134.018302},
	number = {1},
	urldate = {2026-05-26},
	journal = {Physical Review Letters},
	author = {Chao, Xichen and Skipper, Katherine and Royall, C. Patrick and Henkes, Silke and Liverpool, Tanniemola B.},
	month = jan,
	year = {2025},
	pages = {018302},
}

@inproceedings{kaeser_individual_2025,
	address = {Atlanta, GA, USA},
	title = {Individual and {Collective} {Behaviors} in {Soft} {Robot} {Worms} {Inspired} by {Living} {Worm} {Blobs}},
	copyright = {https://doi.org/10.15223/policy-029},
	isbn = {979-8-3315-4139-2},
	url = {https://ieeexplore.ieee.org/document/11127959/},
	doi = {10.1109/ICRA55743.2025.11127959},
	urldate = {2026-05-26},
	booktitle = {2025 {IEEE} {International} {Conference} on {Robotics} and {Automation} ({ICRA})},
	publisher = {IEEE},
	author = {Kaeser, Carina and Kwon, Junghan and Challita, Elio and Tuazon, Harry and Wood, Robert J. and Bhamla, Saad and Werfel, Justin},
	month = may,
	year = {2025},
	pages = {2577--2583},
}

@article{maitra_oriented_2019,
	title = {Oriented {Active} {Solids}},
	volume = {123},
	issn = {0031-9007, 1079-7114},
	url = {https://link.aps.org/doi/10.1103/PhysRevLett.123.238001},
	doi = {10.1103/PhysRevLett.123.238001},
	number = {23},
	urldate = {2026-05-26},
	journal = {Physical Review Letters},
	author = {Maitra, Ananyo and Ramaswamy, Sriram},
	month = dec,
	year = {2019},
	pages = {238001},
}

@article{brauns_active_2026,
	title = {Active {Solids}: {Topological} {Defect} {Self}-{Propulsion} {Without} {Flow}},
	volume = {136},
	issn = {0031-9007, 1079-7114},
	shorttitle = {Active {Solids}},
	url = {https://link.aps.org/doi/10.1103/xv94-xpz2},
	doi = {10.1103/xv94-xpz2},
	number = {5},
	urldate = {2026-05-26},
	journal = {Physical Review Letters},
	author = {Brauns, Fridtjof and O’Leary, Myles and Hernandez, Arthur and Bowick, Mark J. and Marchetti, M. Cristina},
	month = feb,
	year = {2026},
	pages = {058302},
}

@article{gray_propulsion_1955,
	title = {The {Propulsion} of {Sea}-{Urchin} {Spermatozoa}},
	volume = {32},
	copyright = {http://www.biologists.com/user-licence-1-1/},
	issn = {0022-0949, 1477-9145},
	url = {https://journals.biologists.com/jeb/article/32/4/802/12959/The-Propulsion-of-Sea-Urchin-Spermatozoa},
	doi = {10.1242/jeb.32.4.802},
	number = {4},
	urldate = {2026-05-26},
	journal = {Journal of Experimental Biology},
	author = {Gray, J. and Hancock, G. J.},
	month = dec,
	year = {1955},
	pages = {802--814},
}

@article{lighthill_flagellar_1976,
	title = {Flagellar {Hydrodynamics}},
	volume = {18},
	issn = {0036-1445, 1095-7200},
	url = {http://epubs.siam.org/doi/10.1137/1018040},
	doi = {10.1137/1018040},
	number = {2},
	urldate = {2026-05-26},
	journal = {SIAM Review},
	author = {Lighthill, James},
	month = apr,
	year = {1976},
	pages = {161--230},
}

@article{marchetti_hydrodynamics_2013,
	title = {Hydrodynamics of soft active matter},
	volume = {85},
	copyright = {http://link.aps.org/licenses/aps-default-license},
	issn = {0034-6861, 1539-0756},
	url = {https://link.aps.org/doi/10.1103/RevModPhys.85.1143},
	doi = {10.1103/RevModPhys.85.1143},
	number = {3},
	urldate = {2026-05-26},
	journal = {Reviews of Modern Physics},
	author = {Marchetti, M. C. and Joanny, J. F. and Ramaswamy, S. and Liverpool, T. B. and Prost, J. and Rao, Madan and Simha, R. Aditi},
	month = jul,
	year = {2013},
	pages = {1143--1189},
}

@article{ling_instability-driven_2018,
	title = {Instability-driven oscillations of elastic microfilaments},
	volume = {15},
	issn = {1742-5689, 1742-5662},
	url = {https://royalsocietypublishing.org/doi/10.1098/rsif.2018.0594},
	doi = {10.1098/rsif.2018.0594},
	number = {149},
	urldate = {2026-05-26},
	journal = {Journal of The Royal Society Interface},
	author = {Ling, Feng and Guo, Hanliang and Kanso, Eva},
	month = dec,
	year = {2018},
	pages = {20180594},
}

@article{cohen_schrodinger_2023,
	title = {Schrödinger {Dynamics} and {Berry} {Phase} of {Undulatory} {Locomotion}},
	volume = {130},
	issn = {0031-9007, 1079-7114},
	url = {https://link.aps.org/doi/10.1103/PhysRevLett.130.258402},
	doi = {10.1103/PhysRevLett.130.258402},
	number = {25},
	urldate = {2026-05-26},
	journal = {Physical Review Letters},
	author = {Cohen, Alexander E. and Hastewell, Alasdair D. and Pradhan, Sreeparna and Flavell, Steven W. and Dunkel, Jörn},
	month = jun,
	year = {2023},
	pages = {258402},
}

@article{butler_elastohydrodynamics_2026,
	title = {Elastohydrodynamics of three-dimensional chemically active filaments},
	volume = {1029},
	issn = {0022-1120, 1469-7645},
	url = {https://www.cambridge.org/core/product/identifier/S0022112026112087/type/journal_article},
	doi = {10.1017/jfm.2026.11208},
	urldate = {2026-05-26},
	journal = {Journal of Fluid Mechanics},
	author = {Butler, Matthew D. and Walker, Benjamin J. and Montenegro-Johnson, Thomas D. and Katsamba, Panayiota},
	month = feb,
	year = {2026},
	pages = {A42},
}

@article{makanga_instability_2026,
	title = {Instability and self-propulsion of flexible autophoretic filaments},
	volume = {11},
	issn = {2469-990X},
	url = {https://link.aps.org/doi/10.1103/51vg-yb2b},
	doi = {10.1103/51vg-yb2b},
	number = {5},
	urldate = {2026-05-26},
	journal = {Physical Review Fluids},
	author = {Makanga, Ursy and Varma, Akhil and Katsamba, Panayiota},
	month = may,
	year = {2026},
	pages = {053101},
}

@article{hasimoto_soliton_1972,
	title = {A soliton on a vortex filament},
	volume = {51},
	copyright = {https://www.cambridge.org/core/terms},
	issn = {0022-1120, 1469-7645},
	url = {https://www.cambridge.org/core/product/identifier/S0022112072002307/type/journal_article},
	doi = {10.1017/S0022112072002307},
	number = {3},
	urldate = {2026-05-26},
	journal = {Journal of Fluid Mechanics},
	author = {Hasimoto, Hidenori},
	month = feb,
	year = {1972},
	pages = {477--485},
}

@article{de_canio_spontaneous_2017,
	title = {Spontaneous oscillations of elastic filaments induced by molecular motors},
	volume = {14},
	copyright = {http://royalsocietypublishing.org/licence},
	issn = {1742-5689, 1742-5662},
	url = {https://royalsocietypublishing.org/rsif/article/14/136/20170491/64852/Spontaneous-oscillations-of-elastic-filaments},
	doi = {10.1098/rsif.2017.0491},
	number = {136},
	urldate = {2026-05-26},
	journal = {Journal of The Royal Society Interface},
	author = {De Canio, Gabriele and Lauga, Eric and Goldstein, Raymond E.},
	month = nov,
	year = {2017},
	pages = {20170491},
}

@article{clarke_bifurcations_2024,
	title = {Bifurcations and nonlinear dynamics of the follower force model for active filaments},
	volume = {9},
	issn = {2469-990X},
	url = {https://link.aps.org/doi/10.1103/PhysRevFluids.9.073101},
	doi = {10.1103/PhysRevFluids.9.073101},
	number = {7},
	urldate = {2026-05-26},
	journal = {Physical Review Fluids},
	author = {Clarke, Bethany and Hwang, Yongyun and Keaveny, Eric E.},
	month = jul,
	year = {2024},
	pages = {073101},
}

@article{krishnamurthy_emergent_2023,
	title = {Emergent programmable behavior and chaos in dynamically driven active filaments},
	volume = {120},
	issn = {0027-8424, 1091-6490},
	url = {https://pnas.org/doi/10.1073/pnas.2304981120},
	doi = {10.1073/pnas.2304981120},
	number = {28},
	urldate = {2026-05-26},
	journal = {Proceedings of the National Academy of Sciences},
	author = {Krishnamurthy, Deepak and Prakash, Manu},
	month = jul,
	year = {2023},
	pages = {e2304981120},
}

@article{alexander_screw_2025,
	title = {Screw symmetry, chiral hydrodynamics, and odd instability in active cholesterics},
	volume = {112},
	issn = {2470-0045, 2470-0053},
	url = {https://link.aps.org/doi/10.1103/ykzb-pcfx},
	doi = {10.1103/ykzb-pcfx},
	number = {5},
	urldate = {2026-05-28},
	journal = {Physical Review E},
	author = {Alexander, Gareth P. and Kole, S. J. and Maitra, Ananyo and Ramaswamy, Sriram},
	month = nov,
	year = {2025},
	pages = {055424},
}

@article{aditi_simha_hydrodynamic_2002,
	title = {Hydrodynamic {Fluctuations} and {Instabilities} in {Ordered} {Suspensions} of {Self}-{Propelled} {Particles}},
	volume = {89},
	copyright = {http://link.aps.org/licenses/aps-default-license},
	issn = {0031-9007, 1079-7114},
	url = {https://link.aps.org/doi/10.1103/PhysRevLett.89.058101},
	doi = {10.1103/PhysRevLett.89.058101},
	number = {5},
	urldate = {2026-05-28},
	journal = {Physical Review Letters},
	author = {Aditi Simha, R. and Ramaswamy, Sriram},
	month = jul,
	year = {2002},
	pages = {058101},
}

@article{sarkar_mechanochemical_2025,
	title = {Mechanochemical {Feedback} {Drives} {Complex} {Inertial} {Dynamics} in {Active} {Solids}},
	volume = {135},
	issn = {0031-9007, 1079-7114},
	url = {https://link.aps.org/doi/10.1103/19rh-3whq},
	doi = {10.1103/19rh-3whq},
	number = {25},
	urldate = {2026-05-28},
	journal = {Physical Review Letters},
	author = {Sarkar, Siddhartha and Ash, Biswarup and Wu, Yueyang and Boechler, Nicholas and Shankar, Suraj and Mao, Xiaoming},
	month = dec,
	year = {2025},
	pages = {258301},
}

@article{goriely_nonlinear_2000,
	title = {The {Nonlinear} {Dynamics} of {Filaments}},
	volume = {21},
	copyright = {https://www.springernature.com/gp/researchers/text-and-data-mining},
	issn = {0924-090X, 1573-269X},
	url = {https://link.springer.com/10.1023/A:1008366526875},
	doi = {10.1023/A:1008366526875},
	number = {1},
	urldate = {2026-05-28},
	journal = {Nonlinear Dynamics},
	author = {Goriely, Alain and Tabor, Michael},
	month = jan,
	year = {2000},
	pages = {101--133},
}

@book{cross_pattern_2009,
	edition = {1},
	title = {Pattern {Formation} and {Dynamics} in {Nonequilibrium} {Systems}},
	copyright = {https://www.cambridge.org/core/terms},
	isbn = {978-0-521-77935-7 978-0-521-77050-7 978-0-511-62720-0},
	url = {https://www.cambridge.org/core/product/identifier/9780511627200/type/book},
	doi = {10.1017/CBO9780511627200},
	urldate = {2026-05-28},
	publisher = {Cambridge University Press},
	author = {Cross, Michael and Greenside, Henry},
	month = jul,
	year = {2009},
}

@article{lauga_floppy_2007,
	title = {Floppy swimming: {Viscous} locomotion of actuated elastica},
	volume = {75},
	copyright = {http://link.aps.org/licenses/aps-default-license},
	issn = {1539-3755, 1550-2376},
	shorttitle = {Floppy swimming},
	url = {https://link.aps.org/doi/10.1103/PhysRevE.75.041916},
	doi = {10.1103/PhysRevE.75.041916},
	number = {4},
	urldate = {2026-06-08},
	journal = {Physical Review E},
	author = {Lauga, Eric},
	month = apr,
	year = {2007},
	pages = {041916},
}

@article{sartori_curvature_2016,
	title = {Curvature regulation of the ciliary beat through axonemal twist},
	volume = {94},
	copyright = {http://link.aps.org/licenses/aps-default-license},
	issn = {2470-0045, 2470-0053},
	url = {https://link.aps.org/doi/10.1103/PhysRevE.94.042426},
	doi = {10.1103/PhysRevE.94.042426},
	number = {4},
	urldate = {2026-06-08},
	journal = {Physical Review E},
	author = {Sartori, Pablo and Geyer, Veikko F. and Howard, Jonathon and Jülicher, Frank},
	month = oct,
	year = {2016},
	pages = {042426},
}

@book{baker_geometry_2025,
	edition = {1},
	title = {The {Geometry} of {Equilibrium}: {James} {Clerk} {Maxwell} and 21st-{Century} {Structural} {Mechanics}},
	copyright = {https://www.cambridge.org/core/terms},
	isbn = {978-1-009-39764-3 978-1-009-39761-2},
	shorttitle = {The {Geometry} of {Equilibrium}},
	url = {https://www.cambridge.org/core/product/identifier/9781009397643/type/book},
	doi = {10.1017/9781009397643},
	urldate = {2026-06-08},
	publisher = {Cambridge University Press},
	editor = {Baker, William F. and McRobie, Allan},
	month = may,
	year = {2025},
}

@article{blum_biophysics_1979,
	title = {Biophysics of flagellar motility},
	volume = {12},
	copyright = {https://www.cambridge.org/core/terms},
	issn = {0033-5835, 1469-8994},
	url = {https://www.cambridge.org/core/product/identifier/S0033583500002742/type/journal_article},
	doi = {10.1017/S0033583500002742},
	number = {2},
	urldate = {2026-06-08},
	journal = {Quarterly Reviews of Biophysics},
	author = {Blum, Jacob J. and Hines, Michael},
	month = may,
	year = {1979},
	pages = {103--180},
}

@article{warda_elastohydrodynamic_2026,
	title = {Elastohydrodynamic instabilities of a soft robotic arm in a viscous fluid},
	volume = {8},
	issn = {2643-1564},
	url = {https://link.aps.org/doi/10.1103/c914-x8r2},
	doi = {10.1103/c914-x8r2},
	number = {1},
	urldate = {2026-06-08},
	journal = {Physical Review Research},
	author = {Warda, Mohamed and Adhikari, Ronojoy},
	month = mar,
	year = {2026},
	pages = {013229},
}

@article{ashida_non-hermitian_2020,
	title = {Non-{Hermitian} physics},
	volume = {69},
	issn = {0001-8732, 1460-6976},
	url = {https://www.tandfonline.com/doi/full/10.1080/00018732.2021.1876991},
	doi = {10.1080/00018732.2021.1876991},
	number = {3},
	urldate = {2026-06-08},
	journal = {Advances in Physics},
	author = {Ashida, Yuto and Gong, Zongping and Ueda, Masahito},
	month = jul,
	year = {2020},
	pages = {249--435},
}

@article{qiao_control_2022,
	title = {Control of {Active} {Polymeric} {Filaments} by {Chemically} {Powered} {Nanomotors}},
	volume = {18},
	issn = {2331-7019},
	url = {https://link.aps.org/doi/10.1103/PhysRevApplied.18.024051},
	doi = {10.1103/PhysRevApplied.18.024051},
	number = {2},
	urldate = {2026-06-08},
	journal = {Physical Review Applied},
	author = {Qiao, Liyan and Kapral, Raymond},
	month = aug,
	year = {2022},
	pages = {024051},
}

@article{fruchart_non-reciprocal_2021,
	title = {Non-reciprocal phase transitions},
	volume = {592},
	issn = {0028-0836, 1476-4687},
	url = {https://www.nature.com/articles/s41586-021-03375-9},
	doi = {10.1038/s41586-021-03375-9},
	number = {7854},
	urldate = {2026-06-08},
	journal = {Nature},
	author = {Fruchart, Michel and Hanai, Ryo and Littlewood, Peter B. and Vitelli, Vincenzo},
	month = apr,
	year = {2021},
	pages = {363--369},
}

@article{kanso_swimming_2009,
	title = {Swimming due to transverse shape deformations},
	volume = {631},
	copyright = {https://www.cambridge.org/core/terms},
	issn = {0022-1120, 1469-7645},
	url = {https://www.cambridge.org/core/product/identifier/S0022112009006806/type/journal_article},
	doi = {10.1017/S0022112009006806},
	urldate = {2026-06-09},
	journal = {Journal of Fluid Mechanics},
	author = {Kanso, Eva},
	month = jul,
	year = {2009},
	pages = {127--148},
}

@article{sivashinsky_nonlinear_1977,
	title = {Nonlinear analysis of hydrodynamic instability in laminar flames—{I}. {Derivation} of basic equations},
	volume = {4},
	copyright = {https://www.elsevier.com/tdm/userlicense/1.0/},
	issn = {00945765},
	url = {https://linkinghub.elsevier.com/retrieve/pii/0094576577900960},
	doi = {10.1016/0094-5765(77)90096-0},
	number = {11-12},
	urldate = {2026-06-09},
	journal = {Acta Astronautica},
	author = {Sivashinsky, G.I.},
	month = nov,
	year = {1977},
	pages = {1177--1206},
}

@article{kuramoto_persistent_1976,
	title = {Persistent {Propagation} of {Concentration} {Waves} in {Dissipative} {Media} {Far} from {Thermal} {Equilibrium}},
	volume = {55},
	issn = {0033-068X, 1347-4081},
	url = {https://academic.oup.com/ptp/article-lookup/doi/10.1143/PTP.55.356},
	doi = {10.1143/PTP.55.356},
	number = {2},
	urldate = {2026-06-09},
	journal = {Progress of Theoretical Physics},
	author = {Kuramoto, Y. and Tsuzuki, T.},
	month = feb,
	year = {1976},
	pages = {356--369},
}

@incollection{montgomery_gauge_1993,
	title = {Gauge {Theory} of the {Falling} {Cat}},
	url = {http://web.mit.edu/shawest/Public/Papers/cat_gauge_theory.PDF},
	booktitle = {Dynamics and {Control} of {Mechanical} {Systems}},
	publisher = {American Mathematical Society},
	author = {Montgomery, R.},
	editor = {Enos, M.J.},
	year = {1993},
	pages = {193--218},
}

@article{mondal_internal_2020,
	title = {Internal friction controls active ciliary oscillations near the instability threshold},
	volume = {6},
	copyright = {https://creativecommons.org/licenses/by/4.0/},
	issn = {2375-2548},
	url = {https://www.science.org/doi/10.1126/sciadv.abb0503},
	doi = {10.1126/sciadv.abb0503},
	number = {33},
	urldate = {2026-06-09},
	journal = {Science Advances},
	author = {Mondal, Debasmita and Adhikari, Ronojoy and Sharma, Prerna},
	month = aug,
	year = {2020},
	pages = {eabb0503},
}

@article{ranzani_soft_2016,
	title = {A {Soft} {Modular} {Manipulator} for {Minimally} {Invasive} {Surgery}: {Design} and {Characterization} of a {Single} {Module}},
	volume = {32},
	copyright = {https://ieeexplore.ieee.org/Xplorehelp/downloads/license-information/IEEE.html},
	issn = {1552-3098, 1941-0468},
	shorttitle = {A {Soft} {Modular} {Manipulator} for {Minimally} {Invasive} {Surgery}},
	url = {https://ieeexplore.ieee.org/document/7387755/},
	doi = {10.1109/TRO.2015.2507160},
	number = {1},
	urldate = {2026-06-09},
	journal = {IEEE Transactions on Robotics},
	author = {Ranzani, Tommaso and Cianchetti, Matteo and Gerboni, Giada and Falco, Iris De and Menciassi, Arianna},
	month = feb,
	year = {2016},
	pages = {187--200},
}

@article{tekinalp_topology_2024,
	title = {Topology, dynamics, and control of a muscle-architected soft arm},
	volume = {121},
	issn = {0027-8424, 1091-6490},
	url = {https://pnas.org/doi/10.1073/pnas.2318769121},
	doi = {10.1073/pnas.2318769121},
	number = {41},
	urldate = {2026-06-09},
	journal = {Proceedings of the National Academy of Sciences},
	author = {Tekinalp, Arman and Naughton, Noel and Kim, Seung Hyun and Halder, Udit and Gillette, Rhanor and Mehta, Prashant G. and Kier, William and Gazzola, Mattia},
	month = oct,
	year = {2024},
	pages = {e2318769121},
}

@article{elgeti_physics_2015,
	title = {Physics of microswimmers—single particle motion and collective behavior: a review},
	volume = {78},
	issn = {0034-4885, 1361-6633},
	shorttitle = {Physics of microswimmers—single particle motion and collective behavior},
	url = {https://iopscience.iop.org/article/10.1088/0034-4885/78/5/056601},
	doi = {10.1088/0034-4885/78/5/056601},
	number = {5},
	urldate = {2026-06-09},
	journal = {Reports on Progress in Physics},
	author = {Elgeti, J and Winkler, R G and Gompper, G},
	month = may,
	year = {2015},
	pages = {056601},
}

@article{parthasarathy_streaming-enhanced_2019,
	title = {Streaming-enhanced flow-mediated transport},
	volume = {878},
	copyright = {https://www.cambridge.org/core/terms},
	issn = {0022-1120, 1469-7645},
	url = {https://www.cambridge.org/core/product/identifier/S0022112019006438/type/journal_article},
	doi = {10.1017/jfm.2019.643},
	urldate = {2026-06-09},
	journal = {Journal of Fluid Mechanics},
	author = {Parthasarathy, Tejaswin and Chan, Fan Kiat and Gazzola, Mattia},
	month = nov,
	year = {2019},
	pages = {647--662},
}

@article{muller_review_2021,
	title = {Review of the exponential and {Cayley} map on {SE}(3) as relevant for {Lie} group integration of the generalized {Poisson} equation and flexible multibody systems},
	volume = {477},
	issn = {1364-5021, 1471-2946},
	url = {http://royalsocietypublishing.org/rspa/article/82391},
	doi = {10.1098/rspa.2021.0303},
	number = {2253},
	urldate = {2026-06-09},
	journal = {Proceedings of the Royal Society A: Mathematical, Physical and Engineering Sciences},
	author = {Müller, Andreas},
	month = sep,
	year = {2021},
	pages = {20210303},
}

@article{brandenbourger_non-reciprocal_2019,
	title = {Non-reciprocal robotic metamaterials},
	volume = {10},
	issn = {2041-1723},
	url = {https://www.nature.com/articles/s41467-019-12599-3},
	doi = {10.1038/s41467-019-12599-3},
	number = {1},
	urldate = {2026-06-09},
	journal = {Nature Communications},
	author = {Brandenbourger, Martin and Locsin, Xander and Lerner, Edan and Coulais, Corentin},
	month = oct,
	year = {2019},
	pages = {4608},
}

@article{gopal_subramaniam_rigid_2024,
	title = {Rigid flocks, undulatory gaits, and chiral foldamers in a chemically active polymer},
	volume = {26},
	issn = {1367-2630},
	url = {https://iopscience.iop.org/article/10.1088/1367-2630/ad6a7c},
	doi = {10.1088/1367-2630/ad6a7c},
	number = {8},
	urldate = {2026-06-09},
	journal = {New Journal of Physics},
	author = {Gopal Subramaniam, Arvin and Kumar, Manoj and Thutupalli, Shashi and Singh, Rajesh},
	month = aug,
	year = {2024},
	pages = {083009},
}

@misc{sm,
	title={See Supplemental Material at [URL will be inserted by publisher] for animations of rod dynamics and derivations of equations, which includes Ref. [74]}
}

\newpage
\onecolumngrid

\section*{Supplementary Material}
\renewcommand{\theequation}{S\arabic{equation}}
\renewcommand{\thefigure}{S\arabic{figure}}

\section{Geometric flow and shape space equations}

In this section, we give the explicit form of the equations of motion
studied in the Letter. For this, it is useful to introduce the covariant
derivative operator 
\begin{equation}
D=\partial_{u}+\underline{\Pi}\times.
\end{equation}
Combining Eqs. (3) and (4) of the main text, we obtain
\begin{equation}
\begin{aligned}\dot{\underline{h}} & =\frac{1}{\gamma^{T}}D\left(D\underline{F}+\underline{f}\right)+\frac{1}{\gamma^{R}}\underline{h}\times\left(D\underline{M}+\underline{h}\times\underline{F}+\underline{m}\right),\\
\dot{\underline{\Pi}} & =\frac{1}{\gamma^{R}}D\left(D\underline{M}+\underline{h}\times\underline{F}+\underline{m}\right).
\end{aligned}
\end{equation}
We close the equations by substituting the constitutive laws for the
contact forces and moments
\begin{equation}
\underline{F}=\underline{\underline{k}}^{T}(\underline{h}-\underline{e_{1}}),\quad\underline{M}=\underline{\underline{k}}^{R}\underline{\Pi},
\end{equation}
as well as the active force and torque densities
\begin{equation}
\underline{f}=\underline{e_{1}}\times\left(\alpha\underline{\Pi}+\epsilon\underline{\Pi}\times\underline{e_{1}}\right),\quad\underline{m}=\underline{e_{1}}\times\left(\delta\underline{\Pi}+\beta\underline{\Pi}\times\underline{e_{1}}\right).
\end{equation}
We find 
\begin{equation}
\begin{aligned}\dot{\underline{h}} & =\frac{1}{\gamma^{T}}\left(D^{2}\underline{\underline{k}}^{T}[\underline{h}-\underline{e_{1}}]+D\left[\underline{e_{1}}\times(\alpha\underline{\Pi}+\epsilon\underline{\Pi}\times\underline{e_{1}})\right]\right)\\
 & +\frac{1}{\gamma^{R}}\underline{h}\times\left(D\underline{\underline{k}}^{R}\underline{\Pi}+\underline{h}\times\underline{\underline{k}}^{T}[\underline{h}-\underline{e_{1}}]+\underline{e_{1}}\times[\delta\underline{\Pi}+\beta\underline{\Pi}\times\underline{e_{1}}]\right)\\
\dot{\underline{\Pi}} & =\frac{1}{\gamma^{R}}\left(D^{2}\underline{\underline{k}}^{R}\underline{\Pi}+D\left[\underline{h}\times\underline{\underline{k}}^{T}(\underline{h}-\underline{e_{1}})\right]+D\left[\underline{e_{1}}\times(\delta\underline{\Pi}+\beta\underline{\Pi}\times\underline{e_{1}})\right]\right)
\end{aligned}
\end{equation}
The stress-free boundary conditions imply that $\underline{h}=\underline{e_{1}}$
and $\underline{\Pi}=0$ at $u=0,L$ at all times, so that the boundary
conditions are

\begin{equation}
h_{1}=1,h_{2}=h_{3}=0,\quad\Pi_{1}=\Pi_{2}=\Pi_{3}=0
\end{equation}
at both end points of the rod.

\section{Linearized geometric flow on slow manifold}

We linearize the governing equations of our theory, namely Eqs. (3)
and (4) of the main text as follows. The governing equations are closed
by specifying the constitutive laws $\underline{F}=\underline{\underline{k}}^{T}(\underline{h}-\underline{e_{1}})$
and $\underline{M}=\underline{\underline{k}}^{R}\underline{\Pi}$
and the moving frame components of the active force and torque densities
\begin{equation}
\begin{aligned}\underline{f} & =\alpha\begin{bmatrix}0\\
-\Pi_{3}\\
\Pi_{2}
\end{bmatrix}+\epsilon\begin{bmatrix}0\\
\Pi_{2}\\
\Pi_{3}
\end{bmatrix},\\
\underline{m} & =\delta\begin{bmatrix}0\\
-\Pi_{3}\\
\Pi_{2}
\end{bmatrix}+\beta\begin{bmatrix}0\\
\Pi_{2}\\
\Pi_{3}
\end{bmatrix}.
\end{aligned}
\end{equation}
We linearize the equations about the stationary, straight reference
configuration
\begin{equation}
\left(\underline{h}^{(0)},\underline{\Pi}^{(0)}\right)=\left(\underline{e_{1}},\underline{0}\right),\quad\left(\underline{v}^{(0)},\underline{\Omega}^{(0)}\right)=\left(\underline{0},\underline{0}\right).
\end{equation}
The linearized compatibility equations are
\begin{equation}
\dot{\underline{h}}=\underline{v}^{\prime}-\underline{\Omega}\times\underline{e_{1}},\quad\dot{\underline{\Pi}}=\underline{\Omega}^{\prime}.\label{eq:lin_com}
\end{equation}
We kinematically eliminate the shear, stretch, and twist degrees of
freedom by imposing $\underline{h}=0$ and $\Pi_{1}=0$, through which
we obtain $\Omega_{3}=v_{2}^{\prime}$, $\Omega_{2}=-v_{3}^{\prime}$,
and subsequently
\begin{equation}
\dot{\Pi}_{2}=-v_{3}^{\prime\prime},\quad\dot{\Pi}_{3}=v_{2}^{\prime\prime}.\label{eq:linearcompatibility}
\end{equation}
On the other hand, linearizing the force and torque balance in the
filament limit gives
\begin{equation}
\gamma^{T}\underline{v}=\underline{F}^{\prime}+\underline{f},\quad\underline{0}=\underline{M}^{\prime}+\underline{e_{1}}\times\underline{F}+\underline{m}.\label{eq:fil_lin_dyn}
\end{equation}
Substituting the constitutive equations into the linearized torque
balance, differentiating once, and combining the result with the linearized
force balance yields
\begin{equation}
\begin{aligned}\gamma^{T}v_{2} & =-k_{\text{\ensuremath{\perp}}}^{R}\Pi_{3}^{\prime\prime}-\beta\Pi_{3}^{\prime}-\alpha\Pi_{3}-\delta\Pi_{2}^{\prime}+\epsilon\Pi_{2},\\
\gamma^{T}v_{3} & =k_{\text{\ensuremath{\perp}}}^{R}\Pi_{2}^{\prime\prime}+\beta\Pi_{2}^{\prime}+\alpha\Pi_{2}-\delta\Pi_{3}^{\prime}+\epsilon\Pi_{3}.
\end{aligned}
\label{eq:lineardynamics}
\end{equation}
We obtain a closed linear system in terms of the perturbations to
the curvature, $\Pi_{2}$ and $\Pi_{3}$, by differentiating \eqref{eq:fil_lin_dyn}
twice and substituting the resulting equations into \eqref{eq:linearcompatibility}.
Introducing the complex-valued curvature $\Pi\equiv\Pi_{2}+i\Pi_{3}$,
we may compactly write the resulting linear system as
\begin{equation}
\gamma^{T}\dot{\Pi}=-k_{\text{\ensuremath{\perp}}}^{R}\Pi^{\prime\prime\prime\prime}-(\beta+i\delta)\Pi^{\prime\prime\prime}-(\alpha-i\epsilon)\Pi^{\prime\prime}.
\end{equation}
After nondimensionalization, we may write the dimensionless form of
the linear system as
\begin{equation}
\dot{\Pi}=-\Pi''''-\mu\Pi'''-\nu\Pi'',\label{eq:linearization-1}
\end{equation}
and $\mu=\left(\beta+i\delta\right)\times L/k_{\text{\ensuremath{\perp}}}^{R}$,
$\nu=\left(\alpha-i\epsilon\right)\times L^{2}/k_{\text{\ensuremath{\perp}}}^{R}$
are nondimensional complex-valued activity parameters. The free-free
boundary conditions on the fully nonlinear system translate to the
following boundary conditions on the linear system \eqref{eq:linearization-1}.
Imposing the zero torque boundary conditions $M_{2}=M_{3}=0$ at $u=0,L$
implies that $\Pi=0$ at $u=0,L$. On the other hand, using the linearized
equation $0=\underline{M}'+\underline{e_{1}}\times\underline{F}+\underline{m}$,
we find that at $u=0,L$
\begin{equation}
F_{2}=-k_{\perp}^{R}\Pi_{3}',\quad F_{3}=k_{\perp}^{R}\Pi_{2}'.
\end{equation}
Thus, imposing the zero force boundary conditions $F_{2}=F_{3}=0$
at $u=0,L$ implies that $\Pi^{\prime}=0$ at $u=0,L$.

\section{Linear stability analysis}

In this section, we investigate the behavior of the linearized dynamics
\begin{equation}
\dot{\Pi}=-\Pi''''-\mu\Pi'''-\nu\Pi'',\label{eq:lin_dyn}
\end{equation}
in terms of the nondimensionalized complex curvature $\Pi=\Pi_{2}+i\Pi_{3}$,
where $\mu=\left(\beta+i\delta\right)\times L/k_{\perp}^{R}$, $\nu=\left(\alpha-i\epsilon\right)\times L^{2}/k_{\perp}^{R}$
are nondimensional complex-valued activity parameters, with boundary
conditions $\Pi=\Pi'=0$ at $u=0,1$. With a slight abuse of notation,
we shall use $\alpha,\beta,\delta,\epsilon$ for the nondimensionalized
active moduli too.

We are interested in the spectrum of the differential operator
\begin{equation}
\mathcal{L}=-\frac{d^{4}}{du^{4}}-\mu\frac{d^{3}}{du^{3}}-\nu\frac{d^{2}}{du^{2}}
\end{equation}
equipped with the boundary conditions
\begin{equation}
\Pi(0)=\Pi'(0)=\Pi(1)=\Pi'(1)=0.
\end{equation}
We equip the space of complex-valued functions on $[0,1]$ with the
standard $L^{2}$ inner product
\begin{equation}
(f,g)=\int_{0}^{1}f(u)^{*}g(u)\,du.\label{eq:innerproduct}
\end{equation}
With respect to the inner product \eqref{eq:innerproduct}, we find,
after repeated integrations by parts, that
\begin{equation}
(f,\mathcal{L}g)=(\mathcal{L}^{\dagger}f,g)
\end{equation}
where
\begin{equation}
\mathcal{L}^{\dagger}=-\frac{d^{4}}{du^{4}}+\mu^{*}\frac{d^{3}}{du^{3}}-\nu^{*}\frac{d^{2}}{du^{2}},
\end{equation}
from which we conclude the following:
\begin{itemize}
\item The $\textbf{apolar, achiral}$ operator
\begin{equation}
\mathcal{L}_{\mathrm{AP}}^{\mathrm{AC}}=-\frac{d^{4}}{du^{4}}-\alpha\frac{d^{2}}{du^{2}}
\end{equation}
is Hermitian.
\item The $\textbf{apolar, chiral}$ operator
\begin{equation}
\mathcal{L}_{\mathrm{AP}}^{\mathrm{C}}=-\frac{d^{4}}{du^{4}}-i\delta\frac{d^{3}}{du^{3}}
\end{equation}
is Hermitian.
\item The $\textbf{polar, achiral}$ operator
\begin{equation}
\mathcal{L}_{\mathrm{P}}^{\mathrm{AC}}=-\frac{d^{4}}{du^{4}}-\beta\frac{d^{3}}{du^{3}}
\end{equation}
is non-Hermitian.
\item The $\textbf{polar, chiral}$ operator
\begin{equation}
\mathcal{L}_{\mathrm{P}}^{\mathrm{C}}=-\frac{d^{4}}{du^{4}}+i\epsilon\frac{d^{2}}{du^{2}}
\end{equation}
is non-Hermitian.
\end{itemize}
We look for separable solutions of the form $\Pi\left(t,u\right)=\Pi_{n}\left(u\right)e^{\zeta_{n}t}$,
leading to the following eigenvalue problem:
\begin{equation}
\Pi_{n}''''+\mu\Pi_{n}'''+\nu\Pi_{n}''=-\zeta_{n}\Pi_{n}\label{eq:ev_prob}
\end{equation}
with boundary conditions $\Pi_{n}\left(0\right)=\Pi_{n}'\left(0\right)=\Pi_{n}\left(1\right)=\Pi_{n}'\left(1\right)=0$
. Below, we will be interested in values of $\mu,\nu$ where various
eigenvalues $\zeta_{n}$ cross the real axis, as this signals the
$n$th mode becoming unstable. In particular, below we consider an
apolar, achiral rod and an apolar, chiral rod, both of which have
Hermitian linearized dynamics as shown above. Thus, both operators
carry real eigenvalues $\zeta_{n}\in\mathbb{R}$.

\subsection{Apolar, achiral rod}

\begin{figure}
\centering
\includegraphics[width=1\textwidth]{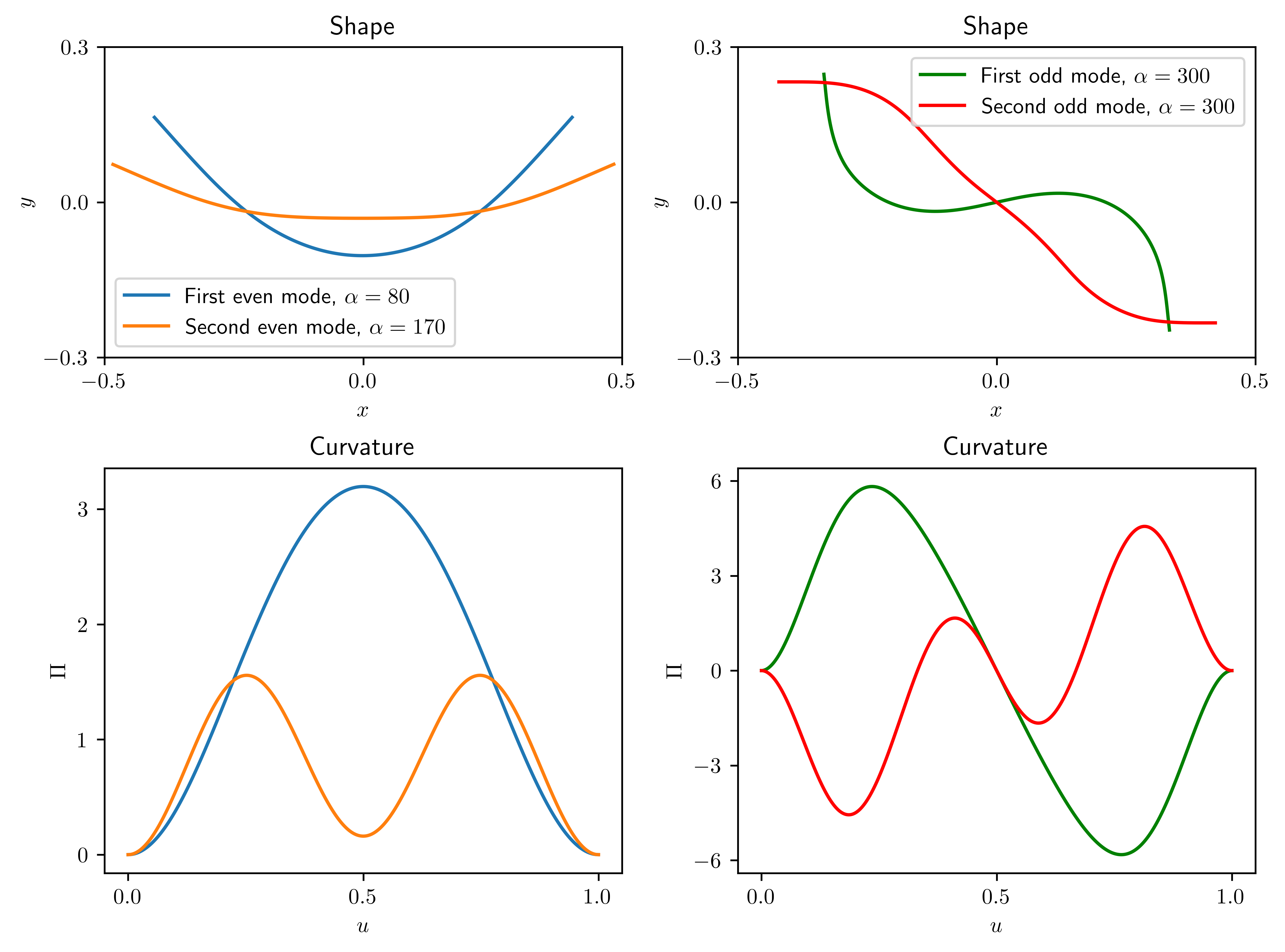}

\caption{Dominant stationary curvature profiles and shapes obtained from linear
stability analysis. Even modes correspond to $\alpha_{n}=4\pi n^{2}$,
while odd modes come from solutions of $\frac{\alpha}{2}=\tan\frac{\alpha}{2}$.
By symmetry, even modes can only translate with a constant velocity,
while odd modes rotate about their geometric center. Shapes were computed
by solving the BVP Eq.~\eqref{eq:bvp_alpha} using the routine \texttt{solve\_bvp}
in \texttt{scipy}.}
\end{figure}
In this subsection, we study a rod which has only $\alpha$ nonzero
and look for positive values of $\alpha$ such that $\zeta_{n}=0$.
(Negative $\alpha$ is an extra line tension and further stabilizes
the straight configuration.) Eq.~\eqref{eq:ev_prob} becomes:
\begin{equation}
\Pi_{n}''''+\alpha\Pi_{n}''=0.\label{eq:bvp_alpha}
\end{equation}
Let $\xi=\sqrt{\alpha}$. The boundary value problem \eqref{eq:bvp_alpha}
has general solution
\[
\Pi_{n}\left(u\right)=A\sin\xi u+B\cos\xi u+Cu+D
\]
for $A,B,C,D$ constants. The boundary conditions lead to the following
system of equations:
\begin{align*}
B+D & =0,\\
\xi A+C & =0,\\
A\sin\xi+B\cos\xi+C+D & =0,\\
A\xi\cos\xi-B\xi\sin\xi+C & =0.
\end{align*}
Eliminating $C,D$ using the first two equations we get $B=-D,C=-\xi A$.
Substituting these into the last two equations we find
\[
\begin{pmatrix}\sin\xi-\xi & \cos\xi-1\\
\xi\left(\cos\xi-1\right) & -\xi\sin\xi
\end{pmatrix}\begin{pmatrix}A\\
B
\end{pmatrix}=\begin{pmatrix}0\\
0
\end{pmatrix}.
\]
This system has a nonzero solution provided that the determinant vanishes,
which, after some algebra, gives
\[
2\xi\sin\frac{\xi}{2}\left(2\sin\frac{\xi}{2}-\xi\cos\frac{\xi}{2}\right)=0.
\]
As $\xi>0$, we get two branches: either $\sin\frac{\xi}{2}=0$, giving
$\xi_{n}=2\pi n$ and $\alpha_{n}=4\pi^{2}n^{2}$ for $n>1$ integer
with a curvature profile $\Pi(u)\propto1-\cos2\pi nu$ corresponding
to an even, self-propelling shape, or $\frac{\xi}{2}=\tan\frac{\xi}{2}$,
which has no closed form solution, but its first nonzero root is roughly
$\xi\approx9$, and the curvature profile is
\[
\Pi(u)\propto\frac{\sin\left(\xi\left(u-\frac{1}{2}\right)\right)}{\cos\frac{\xi}{2}}-\xi\left(u-\frac{1}{2}\right),
\]
corresponding to a rotating $S$-shape. The $U$-shape first appears
at $\alpha_{1}=4\pi^{2}\approx40$, consistently with simulations,
while the $S$-shape only shows up at $\alpha\approx80$, and is typically
metastable.
\begin{figure}
\centering
\includegraphics[width=1\textwidth]{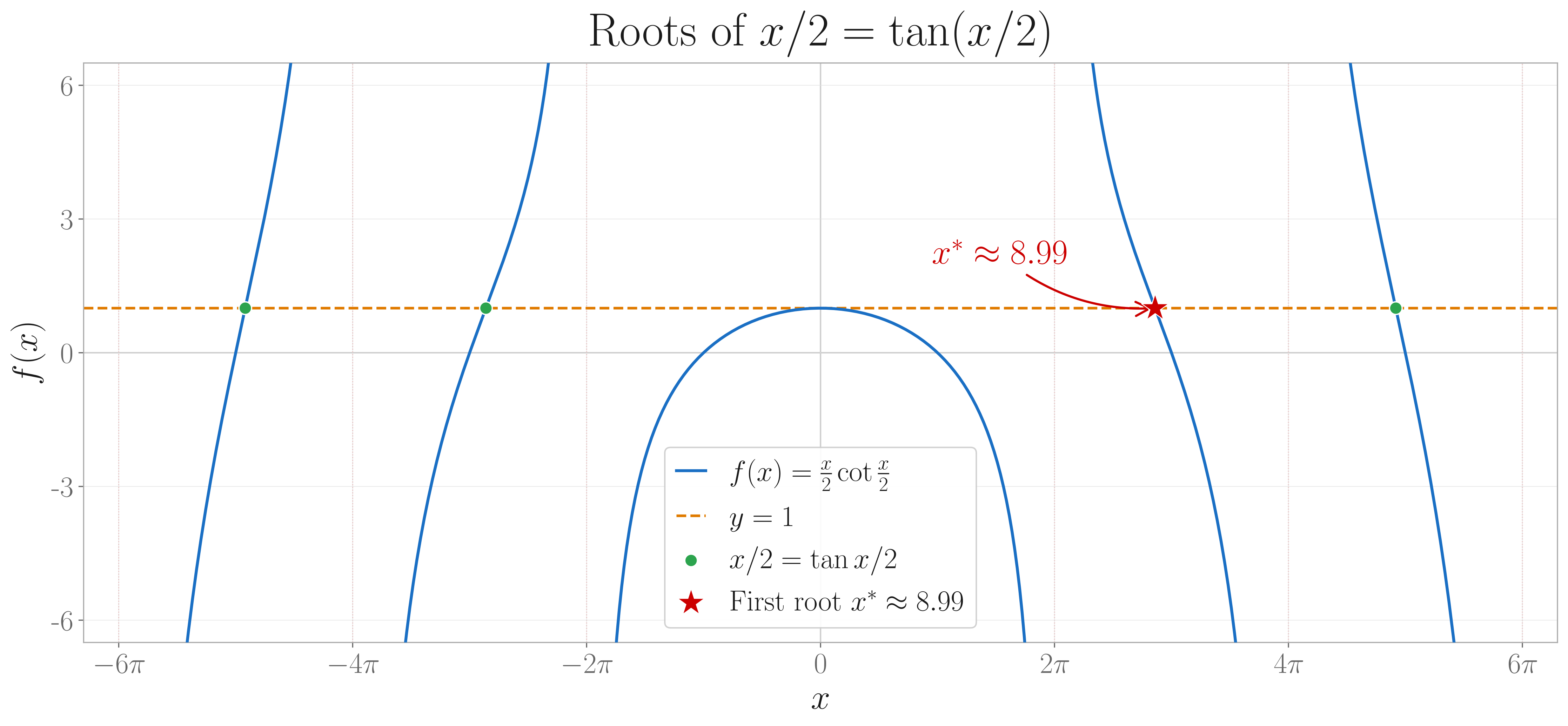}\caption{Roots of the transcendental equation $\frac{\xi}{2}=\tan\frac{\xi}{2}.$}
\end{figure}

\subsection{Apolar, chiral rod.}

In this subsection, we look at a rod which has only $\delta$ nonzero
and look for values of $\delta$ such that $\zeta_{n}=0$. Eq.~\eqref{eq:ev_prob}
becomes:
\begin{equation}
\Pi_{n}''''+i\delta\Pi_{n}'''=0.\label{eq:bvp_delta}
\end{equation}
The boundary value problem \eqref{eq:bvp_delta} has general solution
\begin{equation}
\Pi_{n}\left(u\right)=Ae^{-i\delta u}+Bu^{2}+Cu+D\label{eq:bvp_delta_sol}
\end{equation}
for constants $A,B,C,D$. The boundary conditions lead to the following
system of equations:
\begin{align*}
A+D & =0,\\
-i\delta A+C & =0,\\
Ae^{-i\delta}+B+C+D & =0,\\
-i\delta Ae^{-i\delta}+2B+C & =0.
\end{align*}
Eliminating $C,D$ using the first two equations we get $D=-A,C=i\delta A$.
Substituting these into the last two equations we find
\[
\begin{pmatrix}e^{-i\delta}-1+i\delta & 1\\
-i\delta\left(e^{-i\delta}-1\right) & 2
\end{pmatrix}\begin{pmatrix}A\\
B
\end{pmatrix}=\begin{pmatrix}0\\
0
\end{pmatrix}.
\]
This system has a nonzero solution provided that the determinant vanishes,
which, after some algebra, gives
\[
\frac{\delta}{2}=\tan\frac{\delta}{2}.
\]
The rod becomes unstable at the first nonzero root $\delta=\pm\delta_{c}\approx\pm8.99$
of this transcendental equation, irrespective of the sign of $\delta$.
\begin{figure}
\centering
\includegraphics[width=1\textwidth]{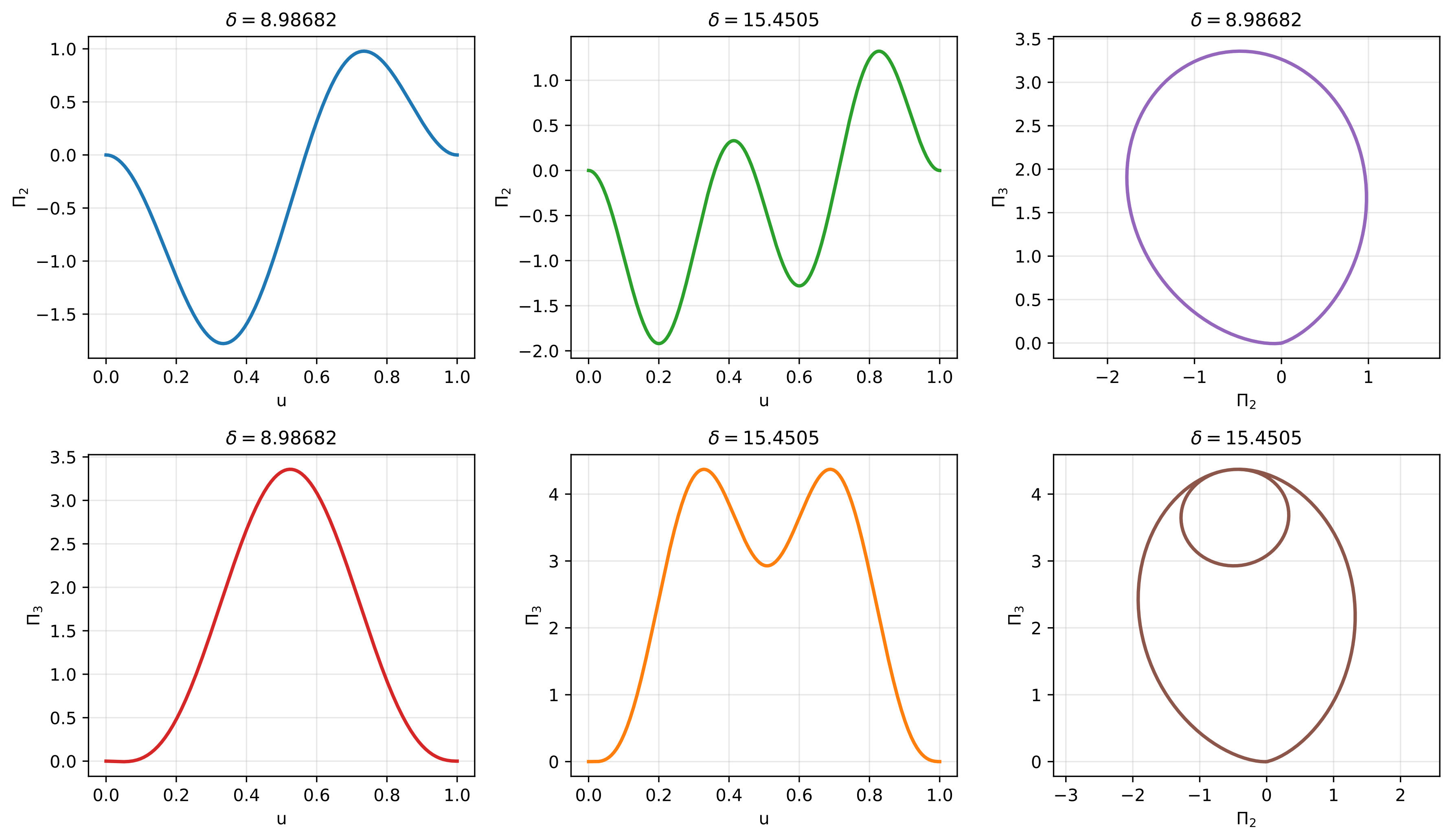}

\caption{First two nonzero solutions to the boundary value problem \eqref{eq:bvp_delta}.}

\end{figure}

\section{Self-propulsion speed at fixed points}

\subsection{Self-propelling U-shape}

We look for translating solutions of the equations of motion, assuming
a planar ($h_{3}=\Pi_{1}=\Pi_{2}=0$) self-propelling U-shape:
\begin{align}
\boldsymbol{v} & =\left(\boldsymbol{F}-\alpha\boldsymbol{e}_{1}\right)'.\label{eq:u_shape_tr}
\end{align}
In Eq. \eqref{eq:u_shape_tr} $\boldsymbol{v}$ is a constant vector.
Integrating \eqref{eq:u_shape_tr} over the spatial extent of the
rod and using the boundary conditions $\boldsymbol{F}=\boldsymbol{0}$
gives:
\begin{equation}
\boldsymbol{v}=-\alpha\left[\boldsymbol{e}_{1}\left(1\right)-\boldsymbol{e}_{1}\left(0\right)\right].\label{eq:sp1}
\end{equation}
Let $\boldsymbol{e}_{1}\left(u\right)=\left(\cos\theta\left(u\right),\sin\theta\left(u\right)\right)$
for a scalar function $\theta\left(u\right)$, which we can take --
by choosing our coordinate system appropriately -- to satisfy $\theta\left(1/2\right)=0$.
Substituting this relation into \eqref{eq:sp1} we find:
\[
\boldsymbol{v}=-\alpha\left(\cos\theta\left(1\right)-\cos\theta\left(0\right),\sin\theta\left(1\right)-\sin\theta\left(0\right)\right).
\]
Since we have an even shape and curvature profile, $\theta$ must
be odd about $u=1/2$, therefore the $x$-component of the self-propulsion
vector is going to be zero, consistently with the symmetry of the
shape. The $y$-component can be simplified as $\theta\left(1\right)=-\theta\left(0\right)$
and $\Pi_{3}=-\theta'$ to find:
\[
\boldsymbol{v}=\alpha\left(0,\sin\left[\int_{1/2}^{1}\Pi_{3}du\right]-\sin\left[\int_{1/2}^{0}\Pi_{3}du\right]\right).
\]
Taking the modulus of both sides, we find the following simple result
for the modulus $v=\left|\boldsymbol{v}\right|$ of the self-propulsion
velocity:
\begin{equation}
v=2\alpha\sin\left(\frac{1}{2}\int_{0}^{1}\Pi_{3}du\right),\label{eq:sp_final_res}
\end{equation}
which is Eq.~(8) of the main text.

\section{Weakly nonlinear analysis}

\subsection{U-shape}

We look for translating solutions of the equations of motion, assuming
a planar shape:
\begin{align*}
\boldsymbol{v} & =\left(\boldsymbol{F}-\alpha\boldsymbol{e}_{1}\right)',\\
0 & =\Pi_{3}'+h_{1}F_{2}-h_{2}F_{1}.
\end{align*}
We work in the filament limit where $h_{1}=1,h_{2}=0$ and $F_{1},F_{2}$
are Lagrange multipliers enforcing these constraints. Differentiating
the first equation once again, we find:
\[
\boldsymbol{F}''=\alpha\boldsymbol{e}_{1}''
\]
Expanding this equation in the moving frame, we get
\[
\left(\underline{F}'+\underline{\Pi}\times\underline{F}\right)'+\underline{\Pi}\times\left(\underline{F}'+\underline{\Pi}\times\underline{F}\right)=\alpha\left(\underline{\Pi}'\times\underline{e_{1}}+\underline{\Pi}\times\left(\text{\underbar{\ensuremath{\Pi}}}\times\underline{e_{1}}\right)\right)
\]
\[
\underline{F}''+2\underline{\Pi}\times\underline{F}'+\underline{\Pi}'\times\underline{F}+\underline{\Pi}\times\left(\underline{\Pi}\times\underline{F}\right)=\alpha\left(\underline{\Pi}'\times\underline{e_{1}}+\underline{\Pi}\times\left(\text{\underbar{\ensuremath{\Pi}}}\times\underline{e_{1}}\right)\right)
\]
The component-wise equations read:

\begin{equation}
\begin{aligned}F_{1}''-2\Pi_{3}F_{2}'-F_{2}\Pi_{3}'+(\alpha-F_{1})\Pi_{3}^{2} & =0,\\
F_{2}''+2\Pi_{3}F_{1}'-F_{2}\Pi_{3}^{2}-(\alpha-F_{1})\Pi_{3}' & =0,\\
\Pi_{3}'+F_{2} & =0.
\end{aligned}
\end{equation}
The boundary conditions are $F_{1}=F_{2}=\Pi_{3}=0$ on both endpoints.
We can eliminate $F_{2}$ using the third equation:
\begin{equation}
\begin{aligned}F_{1}''+2\Pi_{3}\Pi_{3}''+(\Pi_{3}')^{2}+(\alpha-F_{1})\Pi_{3}^{2} & =0\\
\Pi_{3}'''-2\Pi_{3}F_{1}'-\Pi_{3}^{2}\Pi_{3}'+(\alpha-F_{1})\Pi_{3}' & =0
\end{aligned}
\end{equation}
The boundary conditions become $F_{1}=\Pi_{3}=\Pi_{3}'=0$ on both
endpoints. To ease notation, we will drop the indices from $F_{1},\Pi_{3}$
below.

We perform a weakly nonlinear analysis near the instability threshold
$\alpha_{c}=4\pi^{2}$. Let $\alpha=\alpha_{c}+\varepsilon^{2}\Delta$
and expand:
\begin{align*}
F & =\varepsilon F^{(1)}+\varepsilon^{2}F^{(2)}+\varepsilon^{3}F^{(3)}+\dots,\\
\Pi & =\varepsilon\Pi^{(1)}+\varepsilon^{2}\Pi^{(2)}+\varepsilon^{3}\Pi^{(3)}+\dots.
\end{align*}
At linear order we get
\begin{align*}
F^{(1)}{}'' & =0,\\
\Pi^{(1)}{}'''+4\pi^{2}\Pi^{(1)}{}' & =0.
\end{align*}
From the first equation, we get $F^{(1)}=0$, while the second equation
gives $\Pi^{(1)}=A\left(1-\cos\left(2\pi u\right)\right)$.

At quadratic order:
\begin{equation}
\begin{aligned}F^{(2)\prime\prime}+2\Pi^{(1)}\Pi^{(1)\prime\prime}+(\Pi^{(1)\prime})^{2}+4\pi^{2}(\Pi^{(1)})^{2} & =0\\
\Pi^{(2)\prime\prime\prime}+4\pi^{2}\Pi^{(2)\prime} & =0
\end{aligned}
\end{equation}
The first equation can be simplified to:
\begin{equation}
F^{(2)\prime\prime}=4\pi^{2}A^{2}(\cos(4\pi u)-1)
\end{equation}
Integrating it twice and applying the boundary conditions gives:
\begin{equation}
F^{(2)}(u)=A^{2}\left[2\pi^{2}u(1-u)+\frac{1-\cos(4\pi u)}{4}\right]
\end{equation}
At quadratic order, $\Pi^{(2)}\left(u\right)=0$ without loss of generality.

At cubic order, we get:
\begin{align*}
F^{(3)}{}'' & =0,\\
\Pi^{(3)}{}'''+4\pi^{2}\Pi^{(3)}{}' & =2\pi A\left(F^{(2)}-\Delta+A^{2}\left(1-\cos\left(2\pi u\right)\right)^{2}\right)\sin\left(2\pi u\right)+2A\left(1-\cos\left(2\pi u\right)\right)F^{(2)}{}'.
\end{align*}
The amplitude equation for $A$ follows from requiring the forcing
on the RHS of the second equation to have no resonant term:
\[
\int_{0}^{1}\left\{ 2\pi A\left(F^{(2)}-\Delta+A^{2}\left(1-\cos\left(2\pi u\right)\right)^{2}\right)\sin^{2}\left(2\pi u\right)+2A\left(1-\cos\left(2\pi u\right)\right)\sin2\pi uF^{(2)}{}'\right\} du=0.
\]
Carrying out the integral, we find the amplitude equation
\begin{equation}
\left(\frac{35\pi}{8}+\frac{\pi^{3}}{3}\right)A^{3}-\pi\Delta A=0
\end{equation}
The amplitude equation admits the nontrivial steady-state solutions
\begin{equation}
A=\pm\sqrt{\frac{\Delta}{\frac{35}{8}+\frac{\pi^{2}}{3}}}.
\end{equation}
Since the cubic coefficient $\frac{35\pi}{8}+\frac{\pi^{3}}{3}$ is
positive, nonzero solutions exist only for $\Delta>0$, i.e. for ($\alpha>\alpha_{c}$).
The amplitude therefore scales as
\begin{equation}
A\propto\sqrt{\alpha-\alpha_{c}}
\end{equation}
which is the characteristic scaling of a supercritical pitchfork bifurcation.
Thus, the straight configuration loses stability continuously at $\alpha=\alpha_{c}$,
giving rise to a branch of stable U-shaped translating solutions of
arbitrarily small amplitude.

To compute the propulsion speed near the onset of instability, we
use Eq.~\eqref{eq:sp_final_res} to leading order in $\varepsilon$:
\begin{equation}
v\approx2\alpha_{c}\times\frac{A\varepsilon}{2}\int_{0}^{1}\left(1-\cos2\pi u\right)du=C\sqrt{\alpha-\alpha_{c}}
\end{equation}
where
\begin{equation}
C=4\pi^{2}\times\sqrt{\frac{1}{\frac{35}{8}+\frac{\pi^{2}}{3}}}\approx14.676.
\end{equation}

\subsection{Hairpin shape}

We look for rotating solutions of the equations of motion:
\begin{align}
\boldsymbol{0} & =\boldsymbol{F}',\label{eq:hp_st_1}\\
\gamma^{R}\boldsymbol{\omega} & =\boldsymbol{M}'+\boldsymbol{r}'\times\boldsymbol{F}+\boldsymbol{m},\label{eq:hp_st_2}
\end{align}
where $\boldsymbol{\omega}$ is a constant vector. The first equation
can readily be integrated to find $\boldsymbol{F}=\text{const}$,
and from the stress-free boundary conditions we find $\boldsymbol{F}=\boldsymbol{0}$.
Substituting this into the second equation and differentiating in
space, we find
\[
\boldsymbol{0}=\boldsymbol{M}''-\delta\boldsymbol{e}_{1}''.
\]
We expand this equation in the moving frame in terms $\Pi_{2},\Pi_{3}$
and $M_{1}$, which we take to be a Lagrange multiplier enforcing
zero twist $\Pi_{1}=0$. We get:
\[
\left(\underline{M}'+\underline{\Pi}\times\underline{M}\right)'+\underline{\Pi}\times\left(\underline{M}'+\underline{\Pi}\times\underline{M}\right)=\delta\left(\underline{\Pi}'\times\underline{e_{1}}+\underline{\Pi}\times\left(\text{\underbar{\ensuremath{\Pi}}}\times\underline{e_{1}}\right)\right)
\]
\[
\underline{M}''+2\underline{\Pi}\times\underline{M}'+\underline{\Pi}'\times\underline{M}+\underline{\Pi}\times\left(\underline{\Pi}\times\underline{M}\right)=\delta\left(\underline{\Pi}'\times\underline{e_{1}}+\underline{\Pi}\times\left(\text{\underbar{\ensuremath{\Pi}}}\times\underline{e_{1}}\right)\right)
\]
The component-wise equations read:
\[
M_{1}''+\left(\Pi_{2}'\Pi_{3}-\Pi_{3}'\Pi_{2}\right)+\delta\left(\Pi_{2}^{2}+\Pi_{3}^{2}\right)-\left(\Pi_{2}^{2}+\Pi_{3}^{2}\right)M_{1}=0,
\]
\[
\Pi_{2}''-\delta\Pi_{3}'+2\Pi_{3}M_{1}'+\Pi_{3}'M_{1}=0,
\]
\[
\Pi_{3}''+\delta\Pi_{2}'-2\Pi_{2}M_{1}'-\Pi_{2}'M_{1}=0.
\]
These equations are subject to the boundary conditions $\Pi_{2}=\Pi_{3}=M_{1}=0$
at both endpoints, giving rise to a nonlinear boundary value problem.
The straight configuration $M_{1}=\Pi_{2}=\Pi_{3}\equiv0$ is always
a solution, but a nonzero one appears at $\delta=\delta_{c}\approx8.99$
as predicted by the linear stability analysis above.

Linearizing the BVP around the straight configuration we find:
\begin{alignat}{1}
M_{1}'' & =0,\label{eq:hp1}\\
\Pi_{2}''-\delta\Pi_{3}' & =0,\label{eq:hp2}\\
\Pi_{3}''+\delta\Pi_{2}' & =0.\label{eq:hp3}
\end{alignat}
It is easy to see that the linearized system \eqref{eq:hp1}--\eqref{eq:hp3}
has nonzero solutions only if $\delta=2n\pi$ for $n\in\mathbb{Z}$,
meaning that it has no nonzero solution at the onset of the instability
$\delta=\delta_{c}\approx8.99$. What this implies is that there are
no continuously bifurcating solutions of Eqs. \eqref{eq:hp_st_1}--\eqref{eq:hp_st_2}
from the straight configuration near $\delta=\delta_{c}$, therefore
the system must land at the distant fixed point once the straight
configuration has become unstable, explaining the finite amplitude
jump observed in the angular velocity of the hairpin shapes.

\section{Numerical methods}

\subsection{Integration of shape space equations}

The numerical procedure is split into two steps. First, we solve the
closed evolution equations for the strain variables $\left(\underline{h},\underline{\Pi}\right)$
\begin{equation}
\dot{\underline{h}}=\underline{v}'+\underline{\Pi}\times\underline{v}+\underline{h}\times\underline{\Omega},\quad\dot{\underline{\Pi}}=\underline{\Omega}'+\underline{\Pi}\times\underline{\Omega}
\end{equation}
where we eliminate the velocity variables $\left(\underline{v},\underline{\Omega}\right)$
via the force and torque balance
\begin{equation}
\gamma^{T}\underline{v}=\underline{F}'+\underline{\Pi}\times\underline{F}+\underline{f},\quad\gamma^{R}\underline{\Omega}=\underline{M}'+\underline{\Pi}\times\underline{M}+\underline{h}\times\underline{F}+\underline{m},
\end{equation}
combined with the constitutive laws for the stresses
\begin{equation}
\underline{F}=\underline{\underline{k}}^{T}(\underline{h}-\underline{e_{1}}),\quad\underline{M}=\underline{\underline{k}}^{R}\underline{\Pi}
\end{equation}
and the active force and torque densities
\begin{equation}
\underline{f}=-\alpha\underline{\Pi}\times\underline{e_{1}}+\epsilon\underline{e_{1}}\times(\underline{\Pi}\times\underline{e_{1}}),\quad\underline{m}=-\delta\underline{\Pi}\times\underline{e_{1}}+\beta\underline{e_{1}}\times(\underline{\Pi}\times\underline{e_{1}}).
\end{equation}
The closed system of evolution equations for $(\underline{h},\underline{\Pi})$
is first order in time and second order in space. The stress free
boundary conditions are given as
\begin{equation}
\left(\underline{h},\underline{\Pi}\right)=\left(\underline{e_{1}},0\right)\quad\mathrm{at\,\,}u=0,1
\end{equation}
and we consider an initial condition that corresponds to a small perturbation
of the straight undeformed state:
\begin{equation}
\underline{h}(0,u)=\underline{e_{1}}+\varepsilon\sin(2\pi u)(0,1,1),\quad\underline{\Pi}(0,u)=\varepsilon\sin(2\pi u)(0,1,1)
\end{equation}
where $\varepsilon=10^{-3}$. The resulting system of six nonlinear
PDEs for $\left(\underline{h},\underline{\Pi}\right)$ is solved in
Mathematica using the NDSolve function with the method of lines with
adaptive timestepping. We use the following choice of nondimensional
parameters throughout our simulations
\begin{equation}
\frac{k_{\parallel}^{R}}{k_{\perp}^{R}}=\frac{k_{\parallel}^{T}L^{2}}{k_{\perp}^{R}}=\frac{k_{\perp}^{T}L^{2}}{k_{\perp}^{R}}=\frac{\gamma^{T}L^{2}}{\gamma^{R}}=10^{4}.
\end{equation}

\subsection{Reconstruction from shape space trajectory}

The second step of the numerical procedure is to reconstruct the rod
in space and time, given the velocity and strain variables. We may
exploit the $SE(3)=\mathbb{R}^{3}\times SO(3)$ group structure of
the rod kinematics \cite{antman_nonlinear_2004} in Eq.~(1) of the
main text by rewriting it as 
\begin{equation}
\dot{\varphi}=\varphi\mathcal{{V}},\quad\varphi^{\prime}=\varphi\mathcal{{E}}\label{eq:kinematicsabstract}
\end{equation}
where, in block matrix notation, 
\begin{equation}
\varphi=\begin{bmatrix}1 & 0 & 0 & 0\\
\boldsymbol{r} & \boldsymbol{e}_{1} & \boldsymbol{e}_{2} & \boldsymbol{e}_{3}
\end{bmatrix}
\end{equation}
is an $SE(3)-$valued configuration of the rod encoding the centerline
and director data, while
\begin{equation}
\mathcal{{V}}=\begin{bmatrix}0 & 0\\
\underline{v} & \hat{\Omega}
\end{bmatrix}\quad\mathcal{{E}}=\begin{bmatrix}0 & 0\\
\underline{h} & \hat{\Pi}
\end{bmatrix}
\end{equation}
are matrices valued in the Lie algebra $\mathfrak{se}(3)$ of $SE(3)$.
We use the notation $\hat{A}_{ij}=\epsilon_{ikj}A_{k}$ to associate
with each $\underline{A}\in\mathbb{R}^{3}$ an antisymmetric $3\times3$
matrix $\hat{A}$. Given the velocity field $V$ and strain field
$E$ and a spatiotemporal discretization of $[0,T]\times[0,1]$ we
may reconstruct the rod variables in $\varphi$ by making use of the
matrix exponential map and applying a standard Lie-Euler integration
scheme \cite{muller_review_2021} to the kinematic equations Eq.~\eqref{eq:kinematicsabstract}.
We carry out the reconstruction on a uniform grid $t_{n}=n\Delta t,u_{j}=j\Delta u$.
In particular, we carry out a numerical integration of the velocity
equation at $u=0$ via 
\begin{equation}
\varphi\left(t_{n+1},0\right)=\varphi\left(t_{n},0\right)\mathrm{exp}\left(\mathcal{{V}}(t_{n},0)\Delta t\right)
\end{equation}
followed by a numerical integration of the strain equation via
\begin{equation}
\varphi\left(t_{n},u_{j+1}\right)=\varphi\left(t_{n},u_{j}\right)\mathrm{exp}\left(\mathcal{{E}}\left(t_{n},u_{j}\right)\Delta u\right).
\end{equation}
for suitably chosen timestep $\Delta t=2\times10^{-4}$ and spatial
resolution $\Delta u=10^{-2}$.

\subsection{Computation of the phase diagram}

To compute the phase diagram in Fig.~2(g) of the main text, we have
solved the shape space equations using the method outlined above up to $t=0.5$ for various combinations of $\alpha,\beta$ with
$\delta=\epsilon=0$. Points were then classified as limit cycles
if the average time derivative of the curvature of the rod was at
least $1$ at the end of the integration window. Otherwise, if its
average curvature was less than $10^{-4}$, it was classified as a
straight rod, else as a fixed point.

Linear stability boundaries were obtained by discretizing the linear
differential operator on the right-hand side of \eqref{eq:lin_dyn}
using Chebyshev differentiation matrices with $16$ nodes and
boundary conditions $\Pi=0$ and $\Pi'=0$ at endpoints. The first two eigenvalues
of the linear differential operator were then approximated with the
eigenvalues of the corresponding matrix operators.

\section{List of supplementary movies}

\begin{itemize}
    \item Movie 1: Fixed point dynamics
    \item Movie 2: Limit cycle dynamics
    \item Movie 3: Chaotic dynamics
\end{itemize}

\end{document}